\pdfoutput=1

\documentclass[twocolumn,twoside,nofootinbib,showpacs,prd,aps,tightenlines,10pt]{revtex4-1}

\usepackage{amsmath}
\usepackage{graphicx}
\usepackage[hyperfootnotes=true]{hyperref}
\usepackage{xspace}

\newcommand{\eq}[1]{Eq.~\eqref{eq:#1}}
\newcommand{\eqs}[2]{Eqs.~\eqref{eq:#1} and \eqref{eq:#2}}
\renewcommand{\sec}[1]{Sec.~\ref{sec:#1}}

\newcommand{\subsec}[1]{Sec.~\ref{subsec:#1}}

\newcommand{\app}[1]{App.~\ref{app:#1}}
\newcommand{\fig}[1]{Fig.~\ref{fig:#1}}

\newcommand{\ord}[1]{{\mathcal O}(#1)}
\newcommand{\ORd}[1]{{\mathcal O}\Bigl(#1\Bigr)}

\newcommand{\nn}{\nonumber}

\newcommand{\df}{\mathrm{d}}

\newcommand{\tr}{\mathrm{tr}}

\newcommand{\al}{\alpha}
\newcommand{\bt}{\beta}
\newcommand{\ga}{\gamma}
\newcommand{\Ga}{\Gamma}
\newcommand{\de}{\delta}
\newcommand{\eps}{\epsilon}
\newcommand{\si}{\sigma}

\newcommand{\cG}{{\mathcal G}}
\newcommand{\cJ}{{\mathcal J}}

\newcommand{\jet}{\mathrm{jet}}
\newcommand{\fin}{\mathrm{fin}}
\newcommand{\lqcd}{\Lambda_\mathrm{QCD}}

\newcommand{\Pythia}{\textsc{Pythia}\xspace}
\newcommand{\FastJet}{\textsc{FastJet}\xspace}

\allowdisplaybreaks[2]

\begin{document}


\title{Calculating the Charge of a Jet}

\author{Wouter J.~Waalewijn}
\affiliation{Department of Physics, University of California at San Diego, 
La Jolla, CA 92093, U.S.A. \vspace{-0.5ex}}

\begin{abstract}
Jet charge has played an important role in experimentally testing the Parton Model and the Standard Model, and has many potential LHC applications. The energy-weighted charge of a jet is not an infrared-safe quantity, so hadronization must be taken into account. Here we develop the formalism to calculate it, cleanly separating the nonperturbative from the perturbative physics, which we compute at one-loop order. We first study the average and width of the jet charge distribution, for which the nonperturbative input is related to (dihadron) fragmentation functions. In an alternative and novel approach, we consider the full nonperturbative jet charge distribution and calculate its evolution and jet algorithm corrections, which has a natural Monte Carlo-style implementation. Our numerical results are compared to \Pythia and agree well in almost all cases. This calculation can directly be extended to similar track-based observables, such as the total track momentum generated by an energetic parton.
\end{abstract}

\maketitle

\section{Introduction}

Jet charge has played an important role in studying and testing aspects of the Parton Model and of the Standard Model. It was first proposed as a way of measuring the charge of a quark~\cite{Feynman:1972xm}. To reduce the sensitivity to experimental noise, a weighted-definition of jet charge was proposed in Ref.~\cite{Field:1977fa}, 
\begin{align} \label{eq:Qjet}
  Q_\kappa^i  &= \sum_{h \in \text{jet}} z_h^\kappa Q_h
\,,\end{align}
where the parton $i$ initiates the jet, and $z_h = E_h/E$ is the fraction of the jet energy carried by the hadron $h$ with charge $Q_h$. Various choices for $\kappa$ between 0.2 and 1 have been considered at experiments, and we will comment on the optimal choice in this paper. Ref.~\cite{Field:1977fa} also presented a calculation of the jet charge in a recursive, probabilistic model. 

A first experimental application of jet charge was in deeply-inelastic scattering (DIS) experiments~\cite{Erickson:1979wa, Berge:1979qg,Albanese:1984nv}, finding evidence for quarks in the nucleon. Jet charge also played a role in the measurement of the forward-backward asymmetry in $e^+e^-$ collisions~\cite{Stuart:1989db,Decamp:1991se}, which tests the electroweak sector of the Standard Model. Here jet charge was used (roughly speaking) to distinguish the quark and anti-quark jet. Other applications of jet charge include: distinguishing $b$ from $\bar b$ in neutral $B$ meson oscillation measurements~\cite{Buskulic:1992sq}, assigning jets to the hadronically decaying $W^+W^-$ in a measurement of the triple-gauge-boson couplings~\cite{Barate:1997ts}, and excluding an exotic top quark model at the Tevatron~\cite{Abazov:2006vd}.

\begin{figure}[b]
\centering
\includegraphics[width=0.48\textwidth]{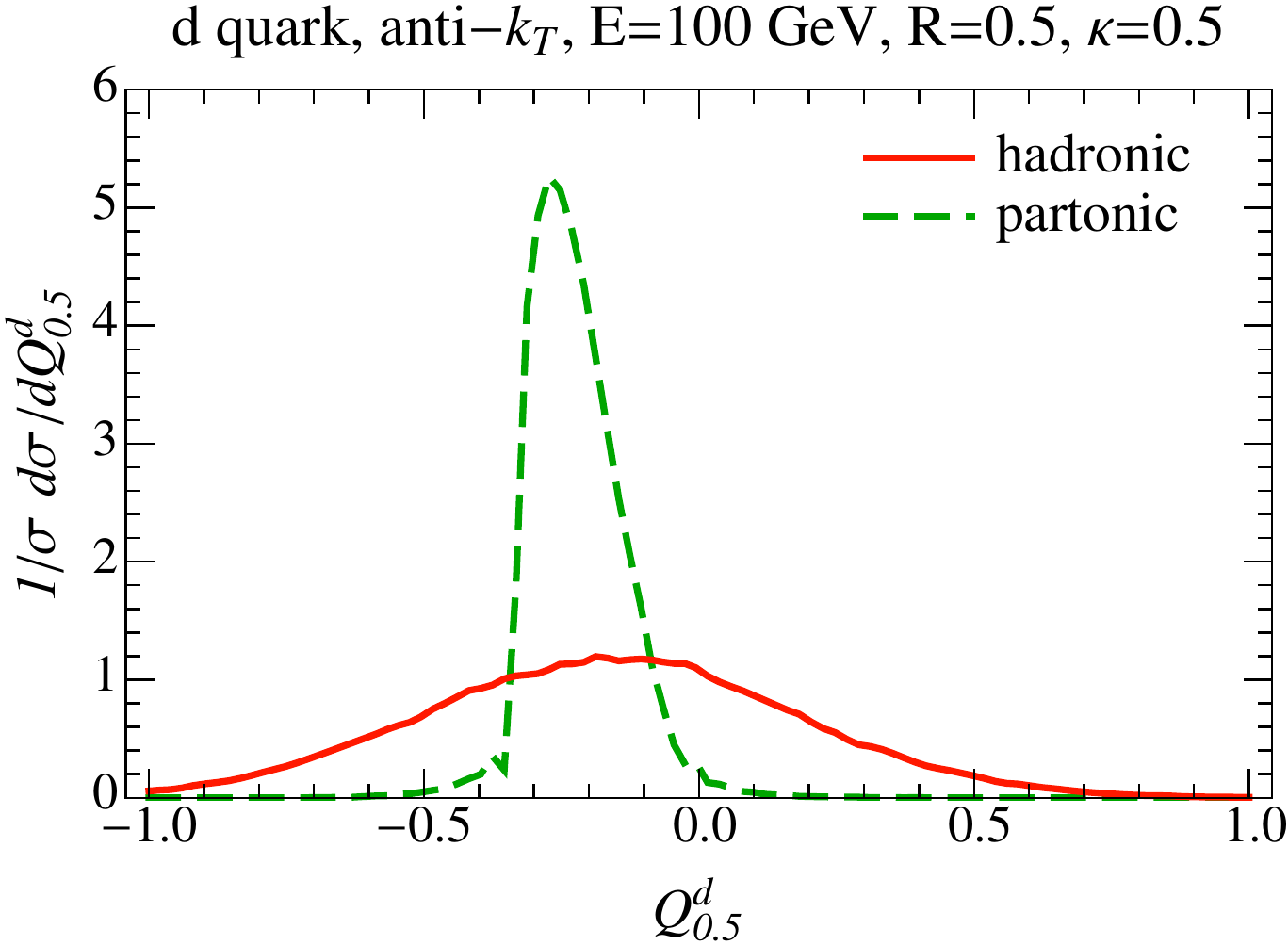}
\caption{The jet charge distribution in \Pythia at the hadronic and partonic level, for a $d$-quark jet with $E = 100$ GeV, using the $e^+ e^-$ anti-$k_T$ algorithm with R=0.5.}
\label{fig:hadr}
\end{figure}

We anticipate that jet charge will play a role at the LHC, since it is almost the only handle we have to separate quark from anti-quark jets or (more ambitiously) distinguish quark flavors~\cite{Krohn:2012fg}. Since jet charge is a track-based observable it is relatively clean from e.g.~pile-up contamination. Calculations of jet charge have so far relied on Monte Carlo programs such as \Pythia~\cite{Sjostrand:2006za,Sjostrand:2007gs}. Here we calculate the jet charge using and extending the framework for fragmentation in jets~\cite{Procura:2009vm,Jain:2011xz,Liu:2010ng,Jain:2011iu,Procura:2011aq,Jain:2012uq} in Soft-Collinear Effective Theory (SCET)~\cite{Bauer:2000ew,Bauer:2000yr,Bauer:2001ct,Bauer:2001yt}. A brief overview of some of our main results was presented in Ref.~\cite{Krohn:2012fg}. An advantage of our approach is a clean separation of the perturbative physics, described by simple analytic formulae, and the nonperturbative physics, which needs to be modeled or extracted from data. In addition, it provides an estimate of the uncertainty and is systematically improvable by calculating higher orders in perturbation theory or including power corrections.

We want to point out that jet charge is not infrared (IR) safe, as the following example illustrates. A jet consisting of a single quark has jet charge $Q_\kappa^q = Q_q$. If it radiates a gluon with energy fraction $z$, it has jet charge $Q_\kappa^q = (1-z)^\kappa Q_q$. However, in the collinear limit these two configurations are indistinguishable. Since $Q_q \neq (1-z)^\kappa Q_q$, this is not IR safe (unless $\kappa=0$). Jet charge must therefore be defined at the level of hadrons, and hadronization effects are crucial. This is also observed in \Pythia, where the hadronization corrections to jet charge are large, as shown in \fig{hadr}. 

In this paper we follow two approaches. First we calculate the average and width of the jet charge distribution.  These are the two most important quantities that determine the power with which jet charge can be used experimentally to separate quarks from anti-quarks and distinguish quark flavors. Fig.~\ref{fig:hadr} illustrates the challenge: the average jet charge is small compared to the width. We find that the average charge of a quark jet is given by a perturbative coefficient, that contains the dependence on the jet algorithm, multiplied by a nonperturbative number, that can be related to fragmentation functions (FFs)~\cite{Collins:1981uk, Collins:1981uw}. The situation is similar, but more complicated, for the width of the jet charge distribution. The width of quark and gluon jets mix and dihadron fragmentation functions now also contribute to the nonperturbative input. 

To go beyond the average and width of the jet charge, we pursue a second approach where we start with a nonperturbative charge distribution. We determine the evolution and the jet algorithm corrections to this distribution at one-loop order. On taking the appropriate moments, this reduces to our first approach. A novel feature of this second approach is that the equations can be naturally solved through a parton shower plus hadronization model.

We will show numerical results for the average and width of the jet charge, and compare with \Pythia. This comparison will be both at the nonperturbative level, where we take FFs as input, and the perturbative level, where we take \Pythia as input and compare the calculable dependence on the jet energy $E$ and size $R$. The uncertainties on charge-separated FFs are fairly large, so the former comparison is not particularly constraining. We find good agreement (except for small $\kappa$), suggesting that \Pythia suffices for initial studies of jet charge. Of course this may change as the precision of FFs or the knowledge of jet charge progresses. 

There are other track-based observables to which our framework can be applied, such as the number of charged hadrons (tracks) in a jet. Here one could again impose a $z^\kappa$-weighting as in \eq{Qjet}, or alternatively a cut on $z$, to remove soft radiation and obtain an experimentally viable quantity. For $\kappa=1$ this corresponds to the total track momentum generated by an energetic parton.

In \sec{avewid}, we calculate the average and width of the jet charge distribution, and discuss the relationship with (dihadron) FFs. We also introduce dihadron fragmenting jet functions here, and discuss some of their properties. The approach involving a nonperturbative jet charge distribution is presented in \sec{nonpert}, and its Monte Carlo implementation is described in \sec{shower}. Numerical results and a comparison with \Pythia are contained in \sec{results}. Here we also discuss the optimal choice for $\kappa$. In \app{pert} and \app{nonpert}, we give the perturbative and nonperturbative coefficients that are relevant for our calculation.

\newpage

\section{Average and Width of the \\ Jet Charge Distribution}
\label{sec:avewid}

\begin{figure*}
\centering
\includegraphics[width=0.48\textwidth]{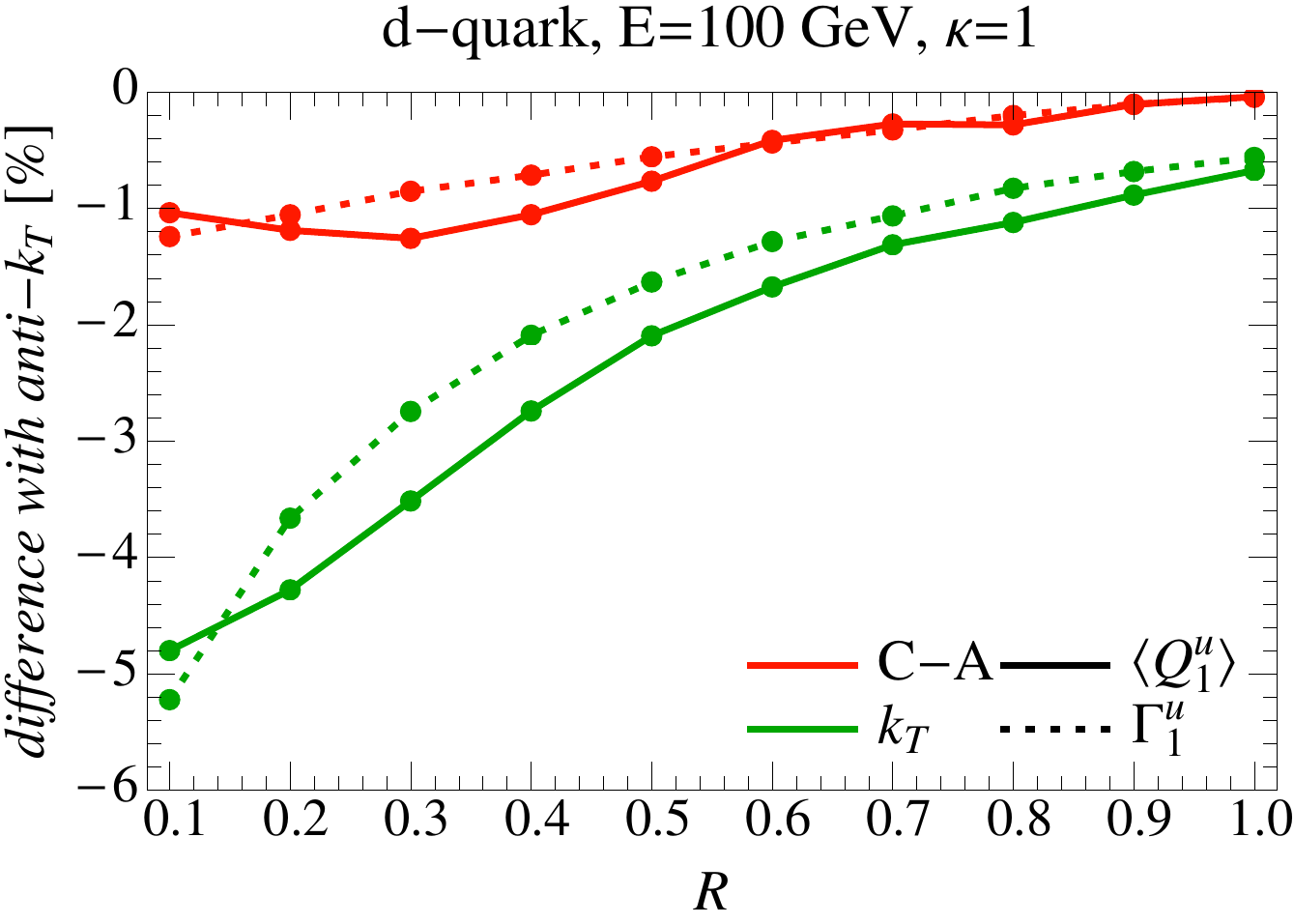} \quad
\includegraphics[width=0.48\textwidth]{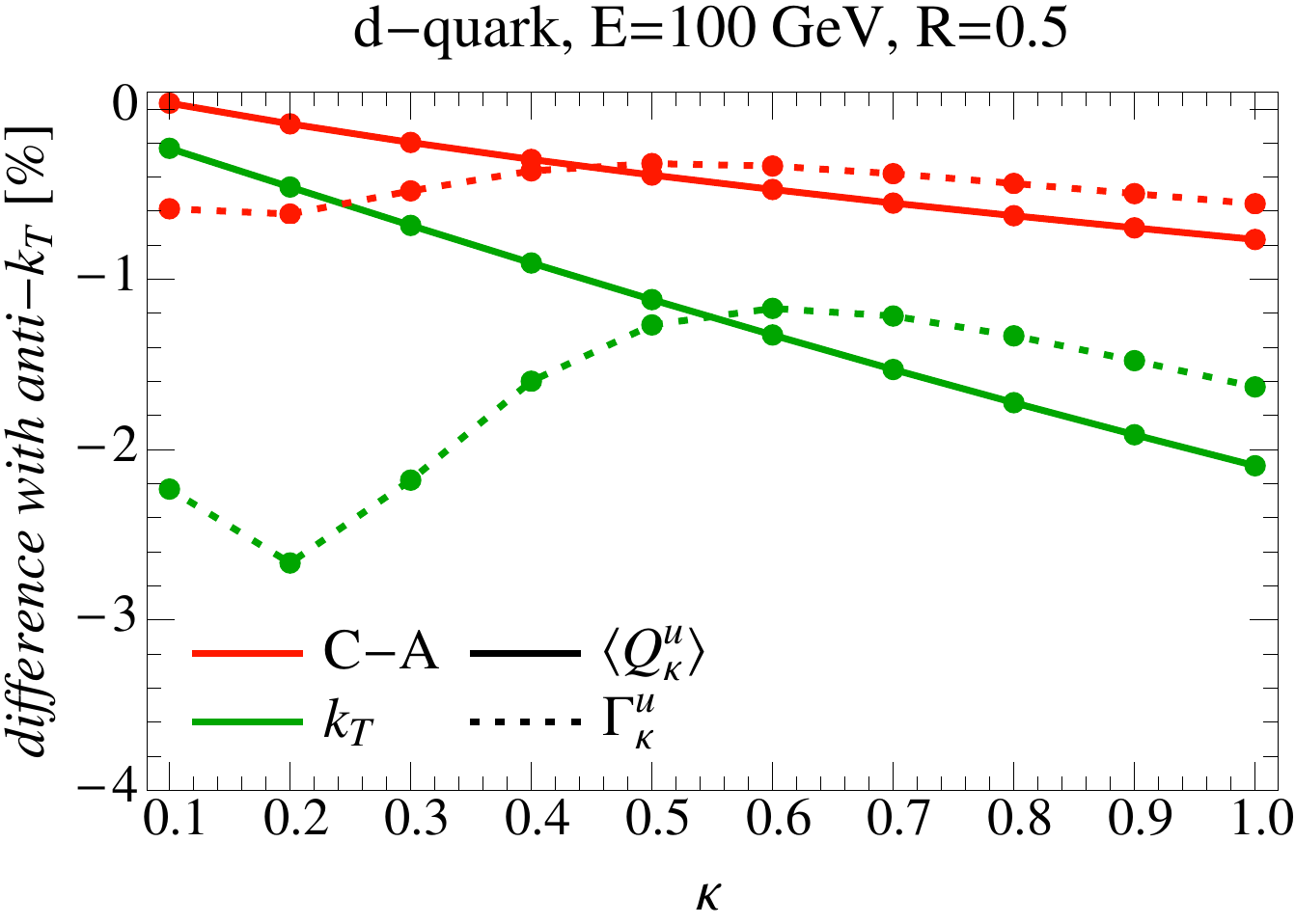} 
\caption{The jet algorithm dependence of the average and width of the jet charge distribution, as function of $\kappa$ and $R$.}
\label{fig:alg}
\end{figure*}

\subsection{Average Jet Charge}
\label{subsec:Qave}

We start out by calculating the average charge for a quark jet. (A gluon jet always has average charge zero.) The average of the jet charge in \eq{Qjet} is 
\begin{align} \label{eq:Qave}
 \langle Q_\kappa^q \rangle 
 & = \frac{1}{\si_{q-\jet}} \int\! \df \si_{q-\jet}\, Q_\kappa(\si_{q-\jet})
 \nn \\ &
  = \int_0^1\! \df z\, z^\kappa \sum_h Q_h\, \frac{1}{\si_{q-\jet}} \frac{\df \si_{h \in q-\jet}}{\df z}
\,,\end{align}
where $\si_{q-\jet}$ is the cross section for producing a quark jet, and $\si_{h \in q-\jet}$ is the cross section for producing the jet in which a hadron $h$ is observed. 

We calculate these cross sections in the framework for jet production of Refs.~\cite{Ellis:2009wj,Ellis:2010rwa,Jouttenus:2009ns}. For simplicity we only consider $e^+e^-$ collisions and use the $e^+e^-$ version of $k_T$-type algorithms, with jet size $R$. (For narrow jets, $R \ll 1$, the extension to $pp$ collisions simply amounts to replacing the jet energy $E$ by the jet transverse momentum $p_T$, as discussed in \app{kt}.) A jet energy veto $\Lambda$ is imposed, to restrict the radiation in the region between jets. The framework is valid for sufficiently narrow, well-separated, energetic jets
\begin{align} \label{eq:pcor}
\tan^2 (R/2),\quad \frac{\tan^2 (R/2)}{\tan^2 (\psi/2)},\quad \frac{\Lambda}{E_\text{min}} \ll 1
\,,\end{align}
where $\psi$ is the minimum angular separation between jets and $E_\text{min}$ the minimum jet energy. It is worth emphasizing that the assumptions in \eq{pcor}, which are used to derive \eq{fact}, are more restrictive than is really necessary. For example, our results will still hold when two jets become close\footnote{The modifications of \eq{fact} for this case can be obtained from Ref.~\cite{Bauer:2011uc}, and do not affect our results.}, as long as these are not the jets whose charge we want to determine.

Using factorization in SCET, the cross sections at leading power are (schematically) given by\footnote{This involves the factorization of the phase-space constraints from the jet algorithm into separate restrictions on the soft and collinear radiation, as discussed in e.g.~Refs.~\cite{Ellis:2010rwa,Walsh:2011fz}.}
\begin{align} \label{eq:fact}
 \si_{q-\jet}\!&= \!\int\! \df \Phi_N \tr[H_N S_N] \Big(\prod_{\ell=1}^{N-1} J_\ell \Big) J_i(E,R,\mu)
\,,  \\ 
  \frac{\df \si_{h\in q-\jet}}{\df z}\!&= \!\int\! \df \Phi_N \tr[H_N S_N] \Big(\prod_{\ell=1}^{N-1} J_\ell \Big) \cG_i^h(E,R,z,\mu) 
\,.\nn \end{align}
The massless $N$-body phase-space for the jets is denoted by $\df \Phi_N$. The hard function $H_N$ describes the hard process, and the soft function $S_N$ the soft radiation. Both $H_N$ and $S_N$ are matrices in color space, and the trace is over color. For each of the jets there is a jet function $J_\ell$ describing the collinear radiation in the jet. To simplify the discussion, we have singled out a jet of flavor $i$ with energy $E$. 

When a hadron is observed in this jet, $J_i$ is replaced by a fragmenting jet function (FJF) $\cG_i^h$~\cite{Procura:2009vm,Jain:2011xz,Procura:2011aq}. We will neglect the contribution from soft radiation to the jet charge, which is suppressed by $\ord{\lambda^{2\kappa}}$, where the size of the SCET power counting parameter $\lambda$ is set by \eq{pcor}. Since soft gluons do not produce an average charge, this additional approximation is absent for the average jet charge. We briefly remind the reader of the most important properties of the FJF: $\cG_i^h$ has the same renormalization as the jet function $J_i$, and can be matched onto fragmentation functions $D_j^h$,
\begin{align} \label{eq:GtoD}
\cG_i^h(E,R,z,\mu) &= \sum_i \int_z^1\! \frac{\df z'}{z'} \cJ_{ij}(E,R,z',\mu) D_j^h\Big(\frac{z}{z'},\mu\Big) 
\nn \\ & \quad \times
\Big[1 + \ORd{\frac{\lqcd^2}{4E^2\tan^2 (R/2)}}\Big]
\,.\end{align}
The perturbative $\cJ_{ij}$ contain the jet algorithm dependence.
The nonperturbative FF $D_j^h$ describes the fragmentation of an energetic hadron $h$ from a parton $j$ in inclusive processes (i.e.~without a jet restriction)~\cite{Collins:1981uk, Collins:1981uw}. 
The FJFs also satisfy certain sum-rules~\cite{Jain:2011xz,Procura:2011aq}.

Inserting \eq{fact} in \eq{Qave}, we find that most pieces cancel in the ratio of cross sections, 
yielding~\cite{Krohn:2012fg}
\begin{align} \label{eq:Qave2}
 \langle Q_\kappa^q \rangle 
  &= \int_0^1\! \df z\, z^\kappa \sum_h Q_h\, \frac{\cG_q^h(E,R,z,\mu)}{2(2\pi)^3 J_q(E,R,\mu)}
  \nn \\
  &= \frac{\widetilde \cJ_{qq}(E,R,\kappa,\mu)}{2(2\pi)^3 J_q(E,R,\mu)} \sum_h Q_h \widetilde D_q^h(\kappa,\mu)
\,.\end{align}
Thus jet charge is independent of the process, up to the power corrections in \eq{pcor}.
This also implies that jet charge is not sensitive to non-global logarithms~\cite{Dasgupta:2001sh,Dasgupta:2002dc} in the soft function~\cite{Kelley:2011ng,Hornig:2011iu,Li:2011zp}.
Because the FJF and jet function have the same anomalous dimension, we see that the $\mu$-dependence cancels in \eq{Qave2}, as should be the case. The last line was obtained using the matching in \eq{GtoD} for the $\kappa$-th moment, with
\begin{align}
  \widetilde \cJ_{ij}(E,R,\kappa,\mu) &= \int_0^1\! \df z\, z^\kappa \cJ_{ij}(E,R,z,\mu)
  \,,\nn \\
  \widetilde D_j^h(\kappa,\mu) &= \int_0^1\! \df z\, z^\kappa D_q^h(z,\mu)
\,.\end{align}
As \eq{Qave2} shows, the average jet charge depends on one nonperturbative number (for each quark flavor and $\kappa$)
\begin{align}
  D_q^Q(\kappa,\mu) = \sum_h Q_h \widetilde D_q^h(\kappa,\mu)
\end{align}
related to FFs. Values for $D_q^Q$ at $\mu=1$ GeV are given in \app{nonpert} for several FF sets. 

The matching coefficients $\cJ_{ij}$ and jet functions $J_i$ that enter in \eq{Qave2} are given at next-to-leading order (NLO) in \app{pert} for $k_T$-type algorithms. (Results for cone algorithms can be found in Ref.~\cite{Procura:2011aq}.) Since there is no distinction between the various $k_T$-like jet algorithms at this order, we study the difference between the average and width of the jet charge distribution in the $e^+e^-$ version of $k_T$~\cite{Catani:1991hj}, Cambridge-Aachen (C-A)~\cite{Dokshitzer:1997in,Wobisch:1998wt} and anti-$k_T$~\cite{Cacciari:2008gp} in \Pythia. The results are shown in \fig{alg}. As these jet algorithms only differ in how they cluster soft radiation, it is not surprising that the difference is only a few \% and grows for small $R$. The dependence on $\kappa$ is counterintuitive: one would expect better agreement for larger values of $\kappa$, since that suppresses the soft radiation. Of course this effect is rather small.

\begin{figure}[t]
\centering
\includegraphics[width=0.48\textwidth]{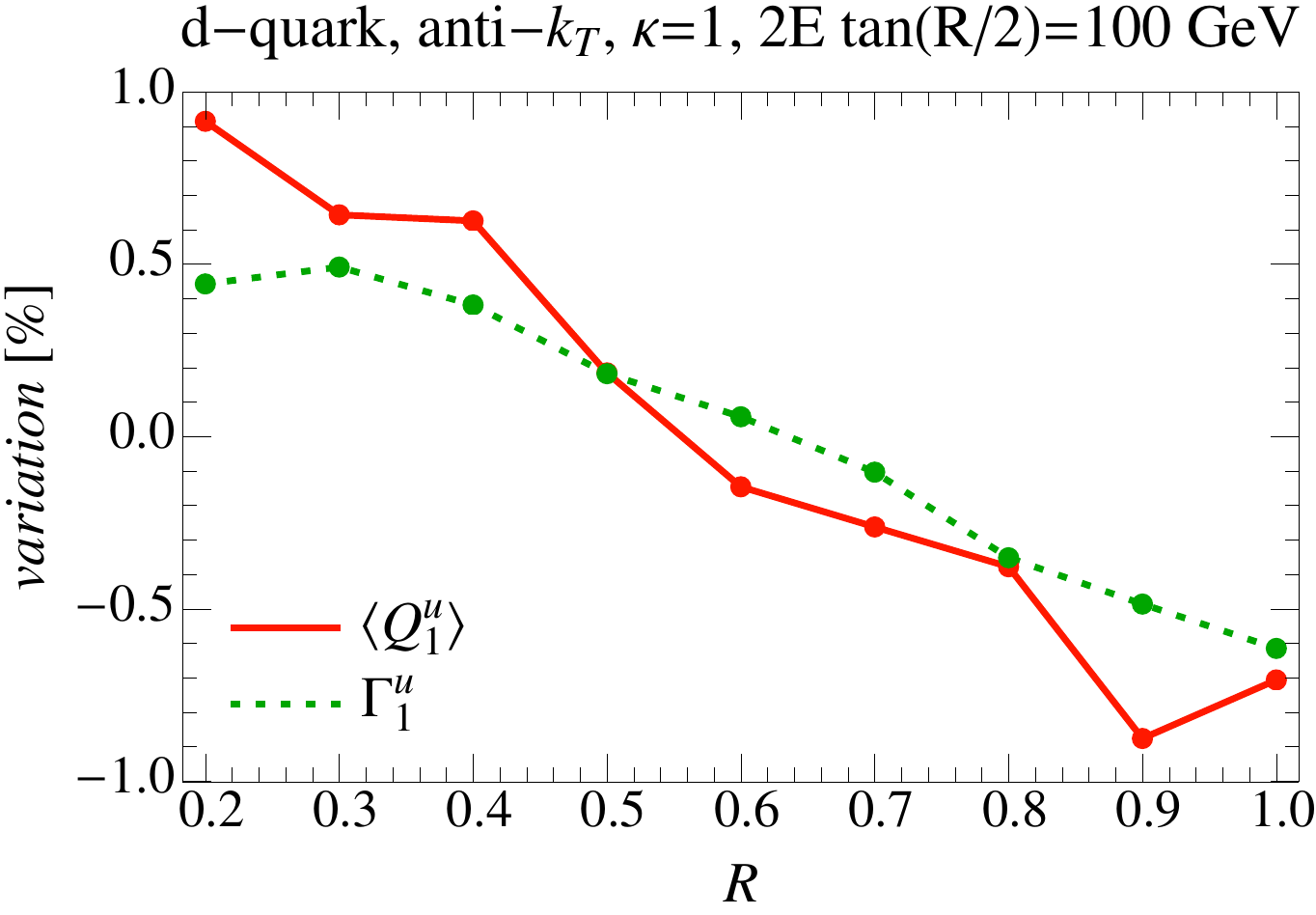}
\caption{The variation of the average and width of the jet charge distribution, keeping $2E \tan(R/2) = 100$ GeV fixed.}
\label{fig:var}
\end{figure}

The matching coefficients $\cJ_{ij}$ and the jet functions $J_i$ contain logarithms of $2 E\tan (R/2)/\mu$, so one should take $\mu \sim 2 E\tan (R/2) \sim ER$ to avoid large logarithms. That this combination of $E$ and $R$ appears can be seen by boosting the jet along the jet axis. This boost invariance is of course spoiled by soft radiation from other jets, which is accounted for by the power corrections in \eq{pcor}. We have investigated to what extent \Pythia shows the same invariance in \fig{var}, which provides an estimate of these power corrections. As you can see, the invariance holds at the percent level\footnote{This is process dependent and is expected to be larger for hadronic collisions or when more jets are present. We have also taken a quite strong cut $\Lambda$ on radiation outside the jets here.}.

To evolve $\widetilde D_q^Q(\kappa,\mu)$ to $\mu \sim 2 E\tan (R/2)$, we need its RGE
\begin{equation} \label{eq:DQ_evo}
  \mu \frac{\df}{\df \mu} \widetilde D_q^Q(\kappa,\mu) = 
 \frac{\al_s(\mu)}{\pi} \widetilde P_{qq}(\kappa) \widetilde D_q^Q(\kappa,\mu)
\,,\end{equation}
which follows directly from that of the FF.
Explicit expressions for the one-loop splitting functions are given in \app{pert}. Note that the mixing with  gluon FFs vanishes to all orders, since a fragmenting gluon produces no net charge. At leading order, the perturbative coefficient in \eq{Qave2} is 1 and the $E$ and $R$ dependence of the jet is simply governed by \eq{DQ_evo}. This shows that the average jet charge reduces (dilutes) for larger values of $E$ and $R$. Of course the relevant question is how this changes relative to the width of the jet charge distribution, which is what we turn to next.

\subsection{Width of the Jet Charge Distribution}

The width of the jet charge distribution is given by
\begin{align} \label{eq:Ga}
(\Ga_\kappa^i)^2 = \langle (Q_\kappa^i)^2 \rangle - \langle Q_\kappa^i \rangle^2
\,.\end{align}
It is the same for quarks and anti-quarks and does not vanish for gluon jets. In the previous section we calculated $\langle Q_\kappa^i \rangle$, so now we will determine
\begin{align}
 \langle (Q_\kappa^i)^2 \rangle 
&=\sum_n \!\sum_{h_1,\ldots,h_n}\! S_{h_1,\ldots,h_n} \int\! \df z_1\cdots \df z_n\, 
\\ &\quad \times
(Q_1 z_1^\kappa\!+\!\cdots\!+\!Q_n z_n^\kappa)^2\, \frac{1}{\si_{i-\jet}} \frac{\df^n \si_{h_1 \cdots h_n \in i-\jet}}{\df z_1\cdots \df z_n} 
\,,\nn\end{align}
where $S_{h_1,\ldots,h_n}$ is the appropriate symmetry factor. After integrating over most of the $z_i$ this simplifies to 
\begin{align} \label{eq:Qsig_def}
  \big\langle (Q_\kappa^i)^2 \big\rangle 
  &= \int_0^1\! \df z\, z^{2\kappa} \sum_h Q_h^2 \frac{1}{\si_{i-\jet}} \frac{\df \si_{h\in i-\jet}}{\df z} +
  \nn \\ 
  & \quad \int_0^1\! \df z_1\! \int_0^1\! \df z_2\, z_1^\kappa z_2^\kappa \!\sum_{h_1,h_2}\! Q_{h_1} Q_{h_2}
  \nn \\
  & \quad \times  \frac{1}{\si_{i-\jet}} \frac{\df \si_{h_1,h_2 \in i-\jet}}{\df z_1\, \df z_2}
\,.\end{align}
Here we used $S_{hh} = 1/2$ and accounted for the factor of 2 for $h_1 \neq h_2$ by including both orderings. The calculation of the first term is completely analogous to that of $\langle Q_\kappa^i \rangle$ in \subsec{Qave}. For the second term we get a dihadron FJF $\cG_i^{h_1 h_2}$ instead of a single-hadron FJF. The relevant properties of the dihadron FJF (discussed in the next section) are that it also has the same renormalization as the jet function $J_i$, and that it can be matched onto (dihadron) fragmentation functions according to \eq{Ghh}. Using this, we find
\begin{align} \label{eq:Qsig}
  & \big\langle(Q_\kappa^i)^2 \big\rangle 
  \nn \\ &\quad
  = \sum_j  \frac{\widetilde \cJ_{ij}(E,R,2\kappa,\mu)}{2(2\pi)^3 J_i(E,R,\mu)} \sum_h Q_h^2 \widetilde D_j^h(2\kappa,\mu) +
  \nn \\ & \qquad
   \int_0^1\! \df z_1\!\int_0^1\! \df z_2\, z_1^\kappa z_2^\kappa \sum_{h_1,h_2} Q_{h_1} Q_{h_2}
  \frac{\cG_i^{h_1 h_2}(E,R,z_1,z_2,\mu)}{2(2\pi)^3 J_i(E,R,\mu)}
  \nn \\ & \quad
  =  \sum_j \Big[ \frac{\widetilde \cJ_{ij}(E,R,2\kappa,\mu)}{2(2\pi)^3 J_i(E,R,\mu)}\widetilde D_j^{Q^2}(2\kappa,\mu) +
  \nn \\ & \qquad
\frac{\widetilde \cJ_{ij}(E,R,2\kappa,\mu)}{2(2\pi)^3 J_i(E,R,\mu)}  \sum_{h_1,h_2} Q_{h_1} Q_{h_2} \widetilde D_j^{h_1 h_2}(\kappa,\kappa,\mu) +
 \nn \\ & \qquad  
 \frac{\widehat \cJ_{ij}(E,R,\kappa,\mu)}{2(2\pi)^3 J_i(E,R,\mu)}  \sum_{h_1,h_2} Q_{h_1} \widetilde D_j^{h_1}(\kappa,\mu) Q_{h_2} \widetilde D_{a(ij)}^{h_2}(\kappa,\mu) \Big]
  \nn \\  & \quad
  =  \sum_j  \frac{\widetilde \cJ_{ij}(E,R,2\kappa,\mu)}{2(2\pi)^3 J_i(E,R,\mu)} \big[\widetilde D_j^{Q^2}(2\kappa,\mu)+\widetilde D_j^{QQ}(\kappa,\mu) \big]
 \nn \\ & \qquad
 - \de_{i,g} \frac{\widehat \cJ_{gq}(E,R,\kappa,\mu)}{(2\pi)^3 J_g(E,R,\mu)} \sum_q \big[\widetilde D_q^Q(\kappa,\mu)\big]^2
\,,\end{align}
as reported in Ref.~\cite{Krohn:2012fg} (except for the last term that only contributes for gluon jets). Here
\begin{align} 
  \widehat \cJ_{ij}(E,R,\kappa,\mu) &= \int_0^1\! \df z\, z^\kappa (1-z)^\kappa \cJ_{ij} (E,R,z,\mu)
\,,\end{align}
and the flavor index $a(ij)$, used in an intermediate step, is given by
\begin{align} \label{eq:aij}
  a(ij) = \left\{
  \begin{tabular}{ll}
   $q$ & \phantom{\ if \ $ij =$} $qg, g \bar q$, \\
   $\bar q$ & \ if \ $ij =$ $\bar q g, g q$, \\
   $g$ & \phantom{\ if \ $ij =$} $gg, qq, \bar q \bar q$,
  \end{tabular}\right.
\end{align}
where the $q$'s and $\bar q$'s are of the same flavor\footnote{This is true at one loop and requires modification at two-loop order where new flavor combinations are possible.}.
The last line in \eq{Qsig} only contributes for $i=g$, and leads to nonlinearities. These are not important due to the smallness of $\widehat \cJ_{gq}$ (see Table~\ref{tab:J_k}), except at very small values of $\mu$ or $\kappa$. 

\eq{Qsig} involves two new nonperturbative numbers
\begin{align}
  \widetilde D_j^{Q^2}(2\kappa,\mu) &= \sum_h Q_h^2 \widetilde D_j^h(2\kappa,\mu)
  \,, \nn \\ 
  \widetilde D_j^{QQ}(\kappa,\mu) &= \sum_{h_1,h_2} Q_{h_1} Q_{h_2} \widetilde D_j^{h_1 h_2}(\kappa,\kappa,\mu) 
\,,\end{align}
where the latter depends on dihadron FFs. To evolve these to the appropriate scale $\mu \sim 2 E\tan (R/2)$, we use
\begin{align} \label{eq:DQQ_evo}
  \mu\, \frac{\df}{\df \mu} \widetilde D_i^{QQ}(\kappa,\mu)
 &= \sum_j \frac{\al_s(\mu)}{\pi} \widetilde P_{ji}(2\kappa) \widetilde D_j^{QQ}(\kappa,\mu)
 \\
 & \quad + \sum_j \frac{\al_s(\mu)}{\pi} \widehat P_{ji}(\kappa)  \widetilde D_j^{Q}(\kappa,\mu) \widetilde D_{a(ij)}^Q(\kappa,\mu)
 \nn \\
 &= \sum_j \frac{\al_s(\mu)}{\pi} \widetilde P_{ji}(2\kappa) \widetilde D_j^{QQ}(\kappa,\mu)
  \nn \\
 & \quad - \de_{i,g} \frac{\al_s(\mu)}{\pi} \widehat P_{qg}(\kappa)  \sum_q \big[\widetilde D_q^{Q}(\kappa,\mu)\big]^2 
\,,\nn\end{align}
which can be obtained by taking moments of the dihadron FF evolution in \eq{Dhh_evo}. The combination $\widetilde D_i^{Q^2}(2\kappa,\mu)+\widetilde D_i^{QQ}(\kappa,\mu)$, which appears in \eq{Qsig}, satisfies this same RG equation. The last line of \eq{DQQ_evo} only contributes for $i=g$, as in \eq{Qsig}. This nonlinear contribution to the evolution is again negligible, unless $\mu$ or $\kappa$ is very small. However, the mixing between quarks and gluons is quite sizable, as we will see in e.g.~\fig{mix}.

\subsection{Dihadron Fragmenting Jet Functions}

The dihadron FJFs $\cG_i^{h_1 h_2}(E,R,z_1,z_2,\mu)$ are a direct extension of the single-hadron FJFs of Refs.~\cite{Procura:2009vm,Jain:2011xz,Procura:2011aq}, where an additional hadron is observed in the jet. This modification of the (IR) state, does not affect the (UV) renormalization, which is therefore the same as the jet function $J_i$.

The dihadron FJFs can be matched onto (dihadron) FFs~\cite{Krohn:2012fg}
\begin{align} \label{eq:Ghh}
& \cG_i^{h_1 h_2}(E,R,z_1,z_2,\mu) 
 \\
&= \sum_j \int\! \frac{\df u}{u^2} \,\cJ_{ij}(E,R,u,\mu) D_j^{h_1 h_2}\Big(\frac{z_1}{u},\frac{z_2}{u},\mu\Big) \nn \\
 & + \sum_{j,k} \int\! \frac{\df u}{u} \frac{\df v}{v} \,\cJ_{ijk}(E,R,u,v,\mu) D_j^{h_1}\Big(\frac{z_1}{u},\mu\Big) D_k^{h_2}\Big(\frac{z_2}{v},\mu\Big)
\nn\,.\end{align}
In the first term the hadrons fragment from the same parton $j$, and the coefficients $\cJ_{ij}$ are the same as in \eq{GtoD}. The nonperturbative dihadron FF $D_j^{h_1 h_2}$ describes the fragmentation of energetic hadrons $h_1$ and $h_2$ from a parton $j$ in inclusive processes (without a jet restriction)~\cite{Konishi:1979cb,Sukhatme:1980vs,Vendramin:1981te}. In the second term of \eq{Ghh} the hadrons fragment from different partons $j$ and $k$, which is described by matching onto two FFs, and does not contribute at tree-level, $\cJ_{ijk}^{(0)} = 0$. Performing the matching by replacing the hadrons by partons, we find at one-loop order that
\begin{align} \label{eq:Jijk}
 \cJ_{ijk}^{(1)}(E,R,u,v,\mu) =  \cJ_{ij}^{(1)}(E,R,u,\mu) \de(1-u-v) \de_{k,a(ij)}
\,.\end{align}
The $\de_{k,a(ij)}$ indicates that at one-loop order the flavor $k$ is completely fixed by $ij$, as described by $a(ij)$ in \eq{aij}.
The dihadron FJFs satisfy the sum rule
\begin{align} \label{eq:Ghhsum}
 &\sum_{h_2} \int\! \df z_2\, z_2\, \cG_i^{h_1 h_2}(E,R,z_1,z_2,\mu) 
 \nn \\ & \quad
 = (1 - z_1) \cG_i^{h_1}(E,R,z_1,z_2,\mu) 
\,,\end{align}
which follows from momentum conservation. 

We now perform some basic checks. First we note that by using \eq{Ghhsum} and the corresponding sum rule for dihadron FFs~\cite{Konishi:1979cb,Vendramin:1981te}
\begin{align}
  \sum_{h_2} \int\! \df z_2\,z_2\, D_j^{h_1h_2}(z_1,z_2,\mu) &= (1\!-\!z_1) D_j^{h_1}(z_1,\mu)
\,,\end{align}
\eq{Ghh} reduces to the matching for the single-hadron FJF in \eq{GtoD}.
Secondly, we verify explicitly at one loop that the anomalous dimension of $\cG_i^{h_1 h_2}$ is equal to that of the jet function $J_i$, by using \eq{Ghh} and \eq{Jijk}. This is a straightforward calculation that also requires the one-loop anomalous dimension~\cite{Procura:2011aq}
\begin{align}
  \mu\, \frac{\df}{\df \mu} \frac{\cJ_{ij}^{(1)}(E,R,u,\mu)}{2(2\pi)^3} &= \ga_{J_i}^{(1)}(E,R,\mu)\, \de(1-u)
  \nn \\ & \quad
  - \frac{\al_s(\mu)}{\pi} P_{ji}(u)
\,,\end{align}
where $\ga_J$ is the jet function anomalous dimension, as well as the RG equation of dihadron FFs~\cite{Konishi:1979cb,Sukhatme:1980vs,Majumder:2004wh}
\begin{align} \label{eq:Dhh_evo}
 & \mu\, \frac{\df}{\df \mu} D_i^{h_1h_2}(z_1,z_2,\mu)
 \\  & 
 = \sum_j \int \frac{\df u}{u^2} \frac{\al_s(\mu)}{\pi} P_{ji}(u) D_j^{h_1h_2}\Big(\frac{z_1}{u},\frac{z_2}{u},\mu\Big)
  \nn \\ & \quad 
 \!+ \!\sum_j \!\int\! \frac{\df u}{u(1\!-\!u)} \frac{\al_s(\mu)}{\pi} P_{ji}(u) D_j^{h_1}\Big(\frac{z_1}{u},\mu\Big) D_{a(ij)}^{h_2}\Big(\frac{z_2}{1\!-\!u},\mu\Big)
\,.\nn\end{align}

\section{A Full Nonperturbative \\ Jet Charge Distribution}
\label{sec:nonpert}

In this section we will take a different approach. Our starting point will be a nonperturbative distribution $D_i(Q,\kappa,\mu)$ for the charge $Q$ of a parton of flavor $i$ for a given $\kappa$. (This is not to be confused with the fragmentation function $D_i^h$, and can be distinguished by its arguments.) We assume that these distributions are normalized,
\begin{align} \label{eq:norm}
  \int\! \df Q\, D_i(Q,\kappa,\mu) = 1
\,.\end{align}
In analogy to \sec{avewid}, we will calculate the RG evolution of $D_i(Q,\kappa,\mu)$ and the corrections from the jet algorithm, which we described by a $\cG_i(E,R,Q,\kappa,\mu)$. 

We start by observing that the one-loop RG evolution consists of splittings $i \to ja(ij)$, with $a(ij)$ given in \eq{aij}. The charge is the sum of the charge of the branches, 
\begin{equation}
Q = z^\kappa Q_1 + (1-z)^\kappa Q_2
\,,\end{equation}
where the rescalings $z^\kappa$ and $(1-z)^\kappa$ are necessary because momentum fractions in the branches are taken with respect to their initiating parton. This suggests the following structure for the renormalization,
\begin{align} \label{eq:Dgen_bare}
D_i^\text{bare}(Q,\kappa,\mu) &= \frac{1}{2} \sum_j \int\! \df Q_1\, \df Q_2\, \df z\, Z^D_{ij}(z,\mu)
  \nn \\ & \quad
  \times D_j(Q_1,\kappa,\mu) D_{a(ij)}(Q_2,\kappa,\mu)  
   \nn \\ & \quad
  \times \de[Q-z^\kappa Q_1-(1-z)^\kappa Q_2]
\,. \end{align}
From the partonic one-loop calculation, we find
\begin{align}
  Z_{ij}^D(z,\mu) = 2\de_{ij} \de(1-z) + \frac{\al_s(\mu)}{2\pi} \frac{1}{\eps} P_{ji}(z)
\,.\end{align}
\eq{Dgen_bare} requires regulating the splitting functions for $z \to 0$, which may be obtained from the familiar $z \to 1$ regularizations, using $P_{gg}(z) = P_{gg}(1-z)$ and $P_{gq}(z) = P_{qq}(1-z)$. Taking the $\mu$-derivative of \eq{Dgen_bare}, we find the following one-loop RG evolution of the charge distribution
\begin{align} \label{eq:Dgen_evo}
  \mu \frac{\df}{\df \mu}\, D_i(Q,\kappa,\mu) &= \frac{1}{2} \sum_j \int\! \df Q_1\, \df Q_2\, \df z\, \ga_{ij}^D(z,\mu)
  \nn \\ & \quad
  \times D_j(Q_1,\kappa,\mu) D_{a(ij)}(Q_2,\kappa,\mu)  
   \nn \\ & \quad
  \times \de[Q-z^\kappa Q_1-(1-z)^\kappa Q_2]
\,, \end{align}
with anomalous dimension
\begin{align}
 \ga_{ij}^D(z,\mu) = \frac{\al_s(\mu)}{\pi} P_{ji}(z) 
\,.\end{align}
A nontrivial property of \eq{Dgen_evo} is that it preserves the normalization in \eq{norm}. Taking the appropriate moments, \eq{Dgen_evo} reduces to the evolution for the average and width of the charge distribution in \eqs{DQ_evo}{DQQ_evo}. One advantage of the approach in this section is that it does not require multihadron FFs for higher integer moments. It also allows us to describe non-integer moments of the jet charge distribution, for which there is no description in terms of multihadron FFs.

The generalization of \eq{Dgen_evo} to $n$-loops is expected to be given by
\begin{align} \label{eq:Dgen_evo_n}
  \mu \frac{\df}{\df \mu}\, D_i(Q,\kappa,\mu) 
  &= \frac{1}{n!}\!\sum_{\{j_k\}}\! \int\! \bigg[\prod_{m=1}^{n+1} \!\!\df Q_m \df z_m\, D_{j_m}\!(Q_m,\kappa,\mu) \bigg] 
   \nn \\ & \quad
  \times \de\Big(1\!-\!\sum_{m=1}^{n+1}\! z_m \Big) \de\Big(Q \!-\! \sum_{m=1}^{n+1} z_m^\kappa Q_m\Big)
 \nn \\ & \quad \times
 \ga_{ij_1 \dots j_{n+1}}^D(z_1,\dots, z_n,\mu)
\,. \end{align}
This becomes increasing nonlinear, but the nonlinearities are of course loop suppressed.

In analogy to the fragmenting jet function, we introduce $\cG_i(E,R,Q,\kappa,\mu)$. This is the \emph{jet} charge distribution, which takes the jet restriction into account. Similar to the renormalization in \eq{Dgen_evo}, we find that the one-loop matching is given by
\begin{align} \label{eq:gnp}
 \cG_i(E,R,Q,\kappa,\mu) &= \frac{1}{2} \sum_j \int\! \df Q_1\, \df Q_2\, \df z\, \cJ_{ij}(E,R,z,\mu) 
  \nn \\ & \quad
  \times D_j(Q_1,\kappa,\mu) D_{a(ij)}(Q_2,\kappa,\mu)  
  \nn \\ & \quad
  \times \de[Q-z^\kappa Q_1-(1-z)^\kappa Q_2]
\,. \end{align}
The matching coefficients $\cJ_{ij}$ are the same as for the FJF, but now also need to be regulated for $z \to 0$. This regularization may be obtained from the $z \to 1$ regularizations, using $\cJ_{gg}(E,R,z,\mu) = \cJ_{gg}(E,R,1-z,\mu)$ and $\cJ_{qg}(E,R,z,\mu) = \cJ_{qq}(E,R,1-z,\mu)$. To ``cancel" the factor of 1/2 in \eq{gnp} at tree-level, we modify the tree-level matching coefficients at $z=0$:
\begin{align}
 \cJ_{qq}^{(0)}(E,R,z,\mu) &= \de(1-z)
 \,,\nn \\
 \cJ_{qg}^{(0)}(E,R,z,\mu) &= \de(z)
 \,,\nn \\
 \cJ_{gg}^{(0)}(E,R,z,\mu) &= \de(z)+\de(1-z)
 \,,\nn \\
 \cJ_{gq}^{(0)}(E,R,z,\mu) &= 0
\,.\end{align}
We have checked that \eq{gnp} preserves the normalization in \eq{norm},
\begin{equation}
  \int\! \df Q\, \cG_i(E,R,Q,\kappa,\mu) = 1
\,.\end{equation}
Taking moments of \eq{gnp},
\begin{align}
\langle Q_\kappa^i \rangle &= \int\! \df Q\, Q\, \cG_i(E,R,Q,\kappa,\mu)
\,, \nn \\
\langle (Q_\kappa^i)^2\rangle  &= \int\! \df Q\, Q^2\, \cG_i(E,R,Q,\kappa,\mu)
\,,\end{align}
results in equations that are consistent with the expressions for the average jet charge and the width of the jet charge distribution in \eqs{Qave2}{Qsig}.

\section{Monte Carlo Implementation}
\label{sec:shower}

An interesting feature of the approach in \sec{nonpert} is that it can be \emph{exactly} recast in terms of a Monte Carlo (MC). Specifically, the solution to the nonlinear differential equation in \eq{Dgen_evo} is naturally obtained using a parton shower. Refs.~\cite{Bauer:2006qp,Bauer:2006mk,Baumgart:2010qf} worked on (improving) parton showers in the context of SCET, by studying the matching and RG evolution of the hard function, which describes the short distance physics. Our approach is orthogonal to this: we are only attempting to describe jet charge (and similar observables) but at higher precision. Our MC consists of a (possible) perturbative splitting at the jet scale $\mu \sim 2E\tan(R/2)$, a parton shower connecting the jet scale and the hadronization scale $\mu_0 \sim 1$ GeV, and a hadronization step at $\mu_0$. We will only keep track of the parton type and momentum fractions in the shower, which are the variables relevant for jet charge, and thus do not attempt to give a fully differential description of the final state.

Starting with a quark or gluon with energy $E$, the first step in generating the jet consists of (the possibility of) a perturbative splitting at the jet scale, corresponding to \eq{gnp}. The splittings are described by the following probability densities,
\begin{align} \label{eq:jprob1}
 P[q \to q(z)g(1\!-\!z)] &= \frac{\cJ_{qq}(E,R,z,\mu)}{2(2\pi)^3 J_q(E,R,\mu)} & 0 \leq z \leq 1\!-\!\delta
\,, \nn \\
 P[g \to g(z)g(1\!-\!z)] &= \frac{\cJ_{gg}(E,R,z,\mu)}{4(2\pi)^3 J_g(E,R,\mu)} & \delta \leq z \leq 1\!-\!\delta
\,, \nn \\
 P[g \to q(z)\bar q(1\!-\!z)] &= \frac{\cJ_{gq}(E,R,z,\mu)}{2(2\pi)^3 J_g(E,R,\mu)} & 0 \leq z \leq 1
\,,\end{align}
where $\mu \sim 2E \tan(R/2)$ to avoid large logarithms. These weights are negative, as is common at NLO, so the phrase ``probability" is not meant in the classical sense. The additional factor of 1/2 for $g \to gg$ is due to identical particles.
Because of the singularities in the $\cJ_{ij}$ (regulated by plus distributions) we introduced a cut-off $\de$. This can be thought of as a resolution parameter, and defines the no-splitting probability,
\begin{align} \label{eq:jprob2}
 P[q \to q] &= \int_{1-\delta}^1\! \df z\, \frac{\cJ_{qq}(E,R,z,\mu)}{2(2\pi)^3 J_q(E,R,\mu)}
\,, \nn \\
 P[g \to g] & = \int_{1-\delta}^1\! \df z\, \frac{\cJ_{gg}(E,R,z,\mu)}{2(2\pi)^3 J_g(E,R,\mu)}
\,.\end{align}
For sufficiently small values of $\delta$ the result becomes independent of $\de$, but smaller values also increases the computation time.

We subsequently carry out the evolution in \eq{Dgen_evo} between the jet scale and the hadronization scale by solving this equation iteratively, with step-size $\df \ln \mu$
\begin{align} \label{eq:evo_it}
  D_i(Q,\kappa, \mu e^{\df \ln \mu}) &= 
  D_i(Q,\kappa, \mu) 
  \\ & \quad
  + \sum_j \int\!\df z\, \frac{\al_s(\mu)}{2\pi} P_{ji}(z)\, \df \ln \mu 
  \nn \\ & \quad
  \times \int\! \df Q_1\, \df Q_2\, \de[Q\!-\!z^\kappa Q_1\!-\!(1\!-\!z)^\kappa Q_2]
  \nn \\ & \quad
  \times D_j(Q_1,\kappa,\mu) D_{a(ij)}(Q_2,\kappa,\mu)  
\,. \nn\end{align}
This builds up a parton shower, where for \emph{each step} $\df \ln \mu$ the splitting probability densities are
\begin{align}
 P[q \to q(z)g(1\!-\!z)] &= \frac{\al_s(\mu)}{\pi}\, P_{qq}(z)\, \df \ln \mu & 0 \leq z \leq 1\!-\!\delta
\,, \nn \\
 P[g \to g(z)g(1\!-\!z)] &= \frac{\al_s(\mu)}{2\pi}\, P_{gg}(z) \df \ln \mu & \delta \leq z \leq 1\!-\!\delta
\,, \nn \\
 P[g \to q(z)\bar q(1\!-\!z)] &= \frac{\al_s(\mu)}{\pi}\, P_{qg}(z) \df \ln \mu & 0 \leq z \leq 1
\,,\end{align}
and the corresponding no-splitting probabilities are
\begin{align}
 P[q \to q] &= 1+ \int_{1-\delta}^1\! \df z\, \frac{\al_s(\mu)}{\pi} P_{qq}(z)\, \df\ln \mu
\,, \nn \\
 P[g \to g] & = 1+ \int_{1-\delta}^1\! \df z\, \frac{\al_s(\mu)}{\pi}\, P_{gg}(z)\, \df\ln \mu
\,.\end{align}
Note that the resolution parameter $\de$ used here can in principle be different from the one in \eqs{jprob1}{jprob2}.

\begin{table}[t]
 \begin{tabular}{|lc|cc|} 
 \hline
$\de$ & $\df \ln \mu$ & $\langle Q_\kappa^q \rangle$ & $\Ga_\kappa^q$ \\ \hline
 0.2 & 0.8 & 0.143 & 0.379 \\ 
 0.2 & 0.4 & 0.142 & 0.374 \\
 0.1 & 0.4 & 0.133 & 0.355 \\ 
 0.05 & 0.2 & 0.128 & 0.342 \\
 0.05 & 0.05 & 0.128 & 0.339 \\ \hline
  \multicolumn{2}{|c|}{analytic} & 0.126 & 0.331 \\
 \hline
 \end{tabular}
 \caption{The average and width of the jet charge distribution obtained from the {\sc JetFrag} MC described in this section, for various values of the parameters $\de$ and $\df \ln \mu$, compared to the analytic calculation. We use the Gaussian toy model described in the text for the nonperturbative input.}
\label{tab:mc}
\end{table}

The description up to this point has been fairly generic and does not rely on the observable. However, now we need to use that in the shower the charge is the $z^\kappa$-weighted sum of the charge of the branches, and that the charge distribution of the branches is sampled over, as described by the last two lines of \eq{evo_it}. We will only sample over the charge distributions of the partons at the end of the shower, randomly assigning them a charge using $D_i(Q,\kappa,\mu_0)$ as a probability distribution. This is our hadronization ``model". It is perhaps surprising that such a simple approach to hadronization is possible, compared to the string fragmentation models used in Monte Carlo programs, but this is because we restrict ourselves to a specific observable. By weighting the charges of these final partons with their momentum fractions, we obtain the jet charge for this event. Generating a sufficient number of events (around $10^5$ to $10^6$ for statistical errors at the percent level) yields a numerical calculation of the full jet charge distribution.

\begin{figure}[t]
\centering
\includegraphics[width=0.48\textwidth]{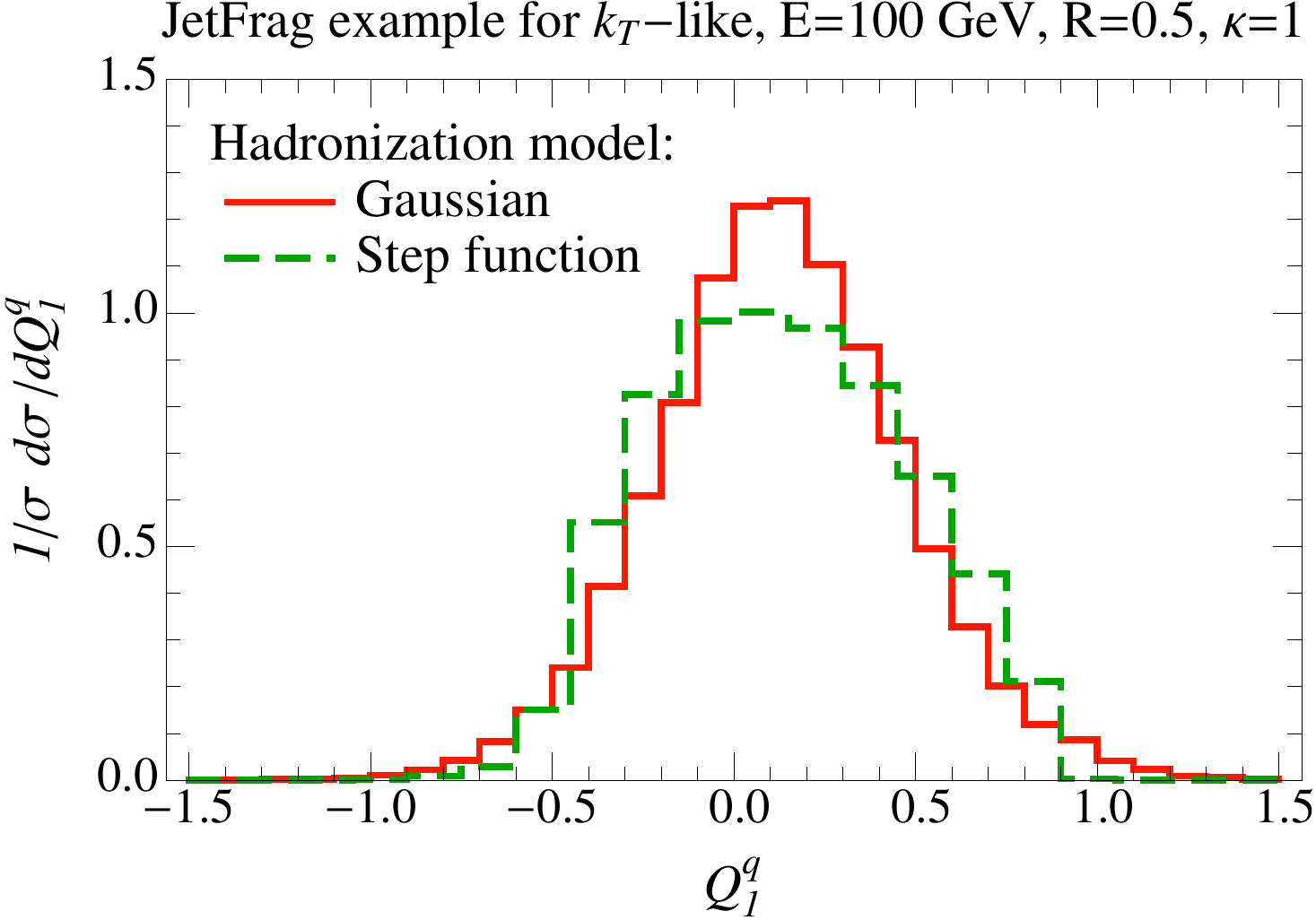} 
\caption{Jet charge distribution obtained using the {\sc JetFrag} Monte Carlo described in \sec{shower}. We use the Gaussian (orange solid) or step function (green dashed) toy model described in the text for the nonperturbative input.}
\label{fig:mc}
\end{figure}

As proof of concept, we have turned this description into a computer program, which we call the {\sc JetFrag} Monte Carlo. This generates unweighted events, but has so far only been minimally optimized. For simplicity we restrict ourselves to $\kappa=1$, and adopt a simple toy model for the nonperturbative input. Specifically, we assume that $D_g(Q,\kappa,\mu_0)$ is Gaussian with average 0 and width 0.6 at the scale $\mu_0 =1$ GeV. Similarly we take $D_q(Q,\kappa,\mu_0) = D_{\bar q}(-Q,\kappa,\mu_0)$ equal for all five\footnote{We here ignore the $b$ and $c$-quark thresholds.} quark flavors, and describe it by a Gaussian with mean 0.2 and width 0.4. In Table~\ref{tab:mc}, we show results for the average and the width of the jet charge distribution, obtained using this MC program for various values of the parameters $\de$ and $\df \ln \mu$. We compare this to the analytic results, to get an idea of how small these parameters need to be for a reliable description. 

The histogram for the corresponding jet charge distribution is shown in \fig{mc} for $\de = \df \ln \mu = 0.05$. Since we start with nonperturbative input that is Gaussian, it is not surprising that the resulting jet charge distribution looks very Gaussian again. As an illustration we therefore also show the result using a step function for $D_g(Q,\kappa,\mu_0)$ and $D_q(Q,\kappa,\mu_0)$ with the same averages and widths. From \sec{avewid} we know that the resulting jet charge distribution will have the same average and width. However, the shape of the distribution is quite different, as shown in \fig{mc}. The parton shower does turn the step function into a more Gaussian distribution, which can presumably be understood as a consequence of the central limit theorem. To the extent that a Gaussian description of jet charge suffices, the analytical approach of \sec{avewid} is of course much simpler.

The procedure described in this section can directly be extended to observables similar to jet charge, by replacing $D_i(Q,\kappa,\mu_0)$ with the appropriate function for that observable. There are no obvious obstacles in extending this approach to  $n$-loop order, although this will involve $1 \to n+1$ splittings [see \eq{Dgen_evo_n}]. It will be interesting to see if a similar approach can be employed for more general track-based jet observables that are also sensitive to soft radiation, such as the track mass of a jet. We leave this question for future work~\cite{Chang:2013rca}. 

\begin{figure}[b]
\centering
\includegraphics[width=0.48\textwidth]{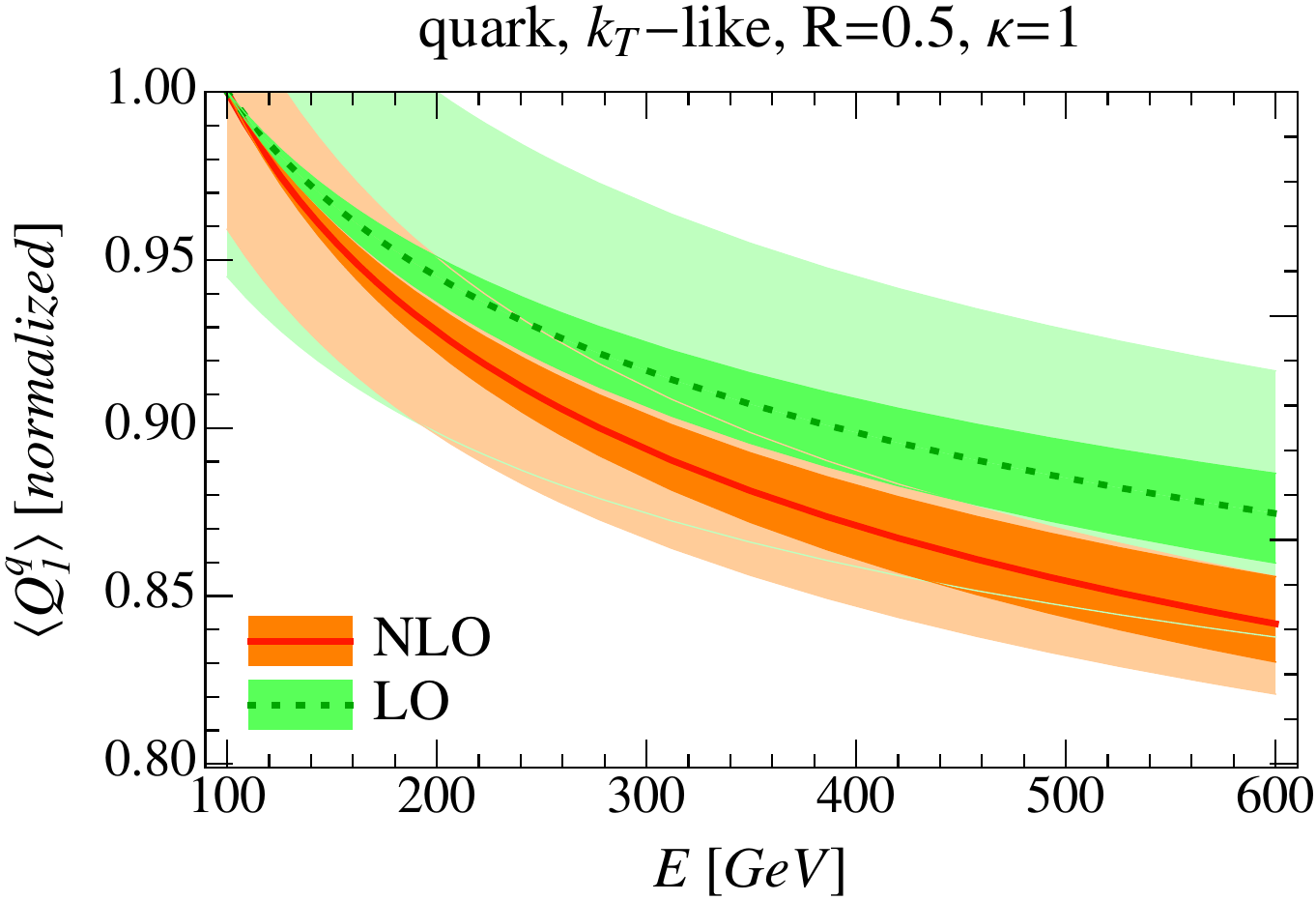} 
\caption{The average charge $\langle Q_1^q \rangle$ at LO and NLO for a $k_T$-like quark jet with R=0.5 and $\kappa=1$. The bands correspond to the perturbative uncertainties as explained in the text.}
\label{fig:aveconv}
\end{figure}

\begin{figure*}[t]
\centering
\includegraphics[width=0.48\textwidth]{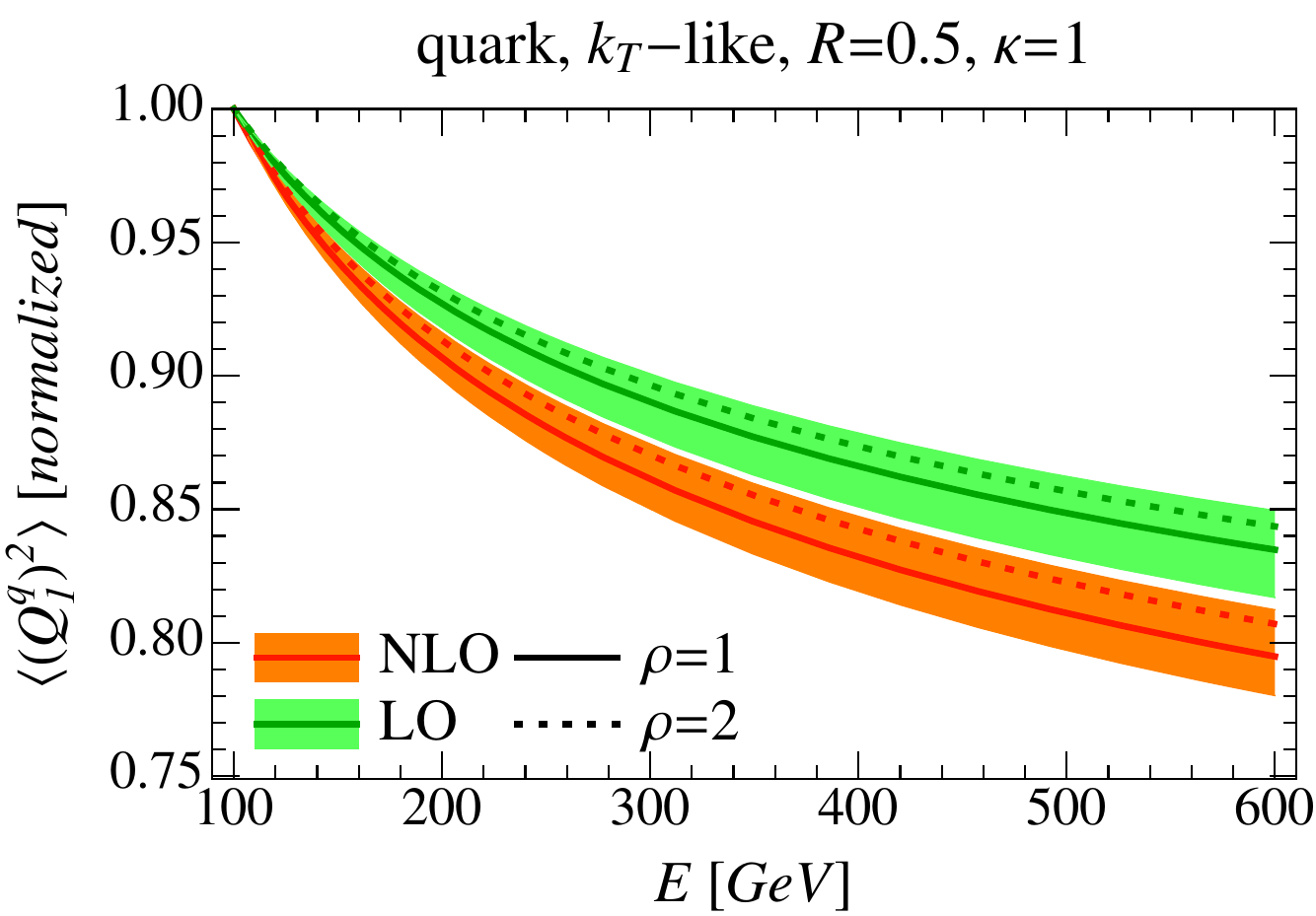} \
\includegraphics[width=0.48\textwidth]{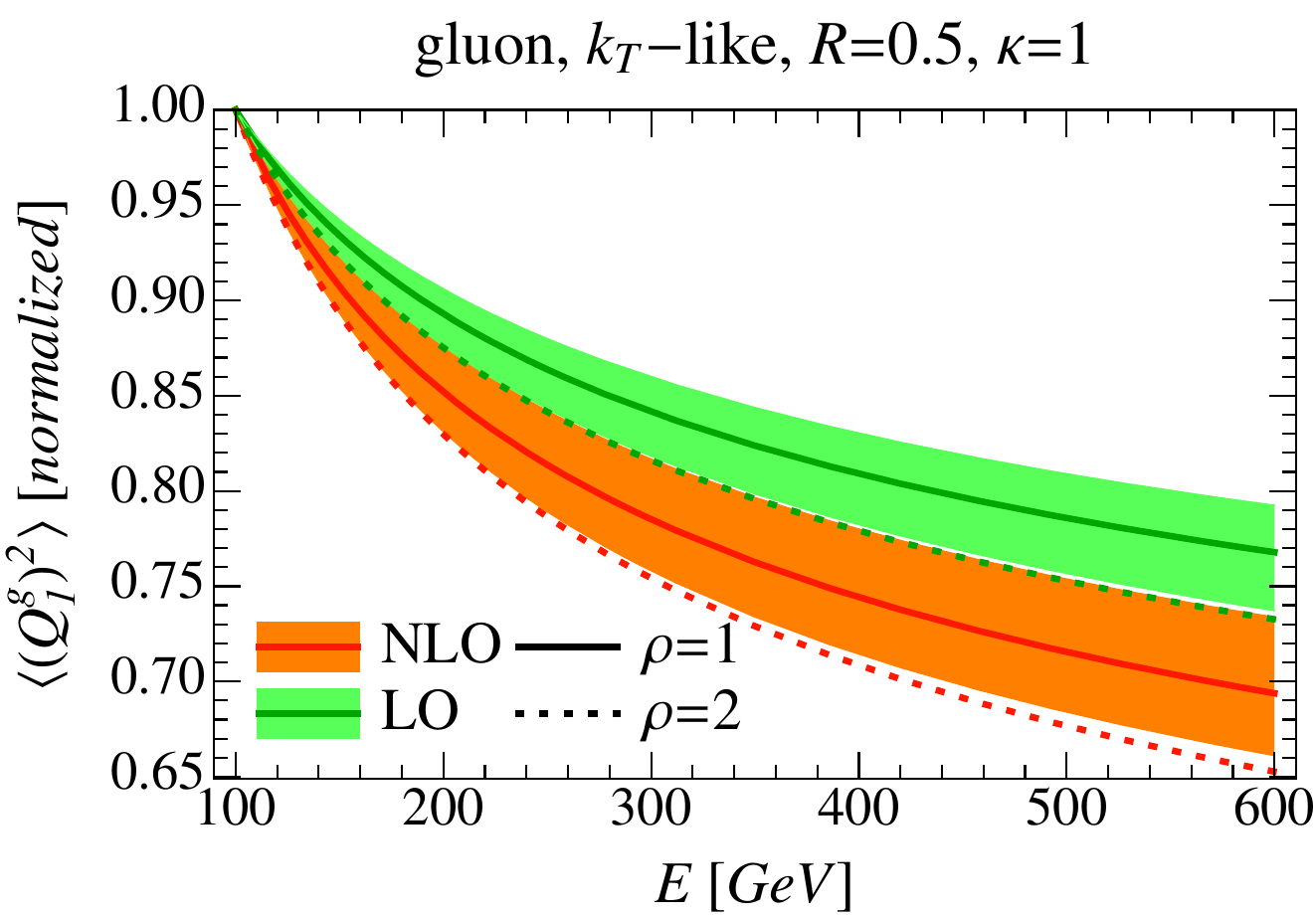}
\caption{$\langle (Q_1^i)^2 \rangle$ at LO and NLO for $k_T$-like quark jets (left panel) and gluon jets (right panel) with R=0.5 and $\kappa=1$. The bands correspond to the perturbative uncertainties for $\rho=1$.}
\label{fig:widconv}
\end{figure*}

\begin{figure*}[t]
\centering
\includegraphics[width=0.46\textwidth]{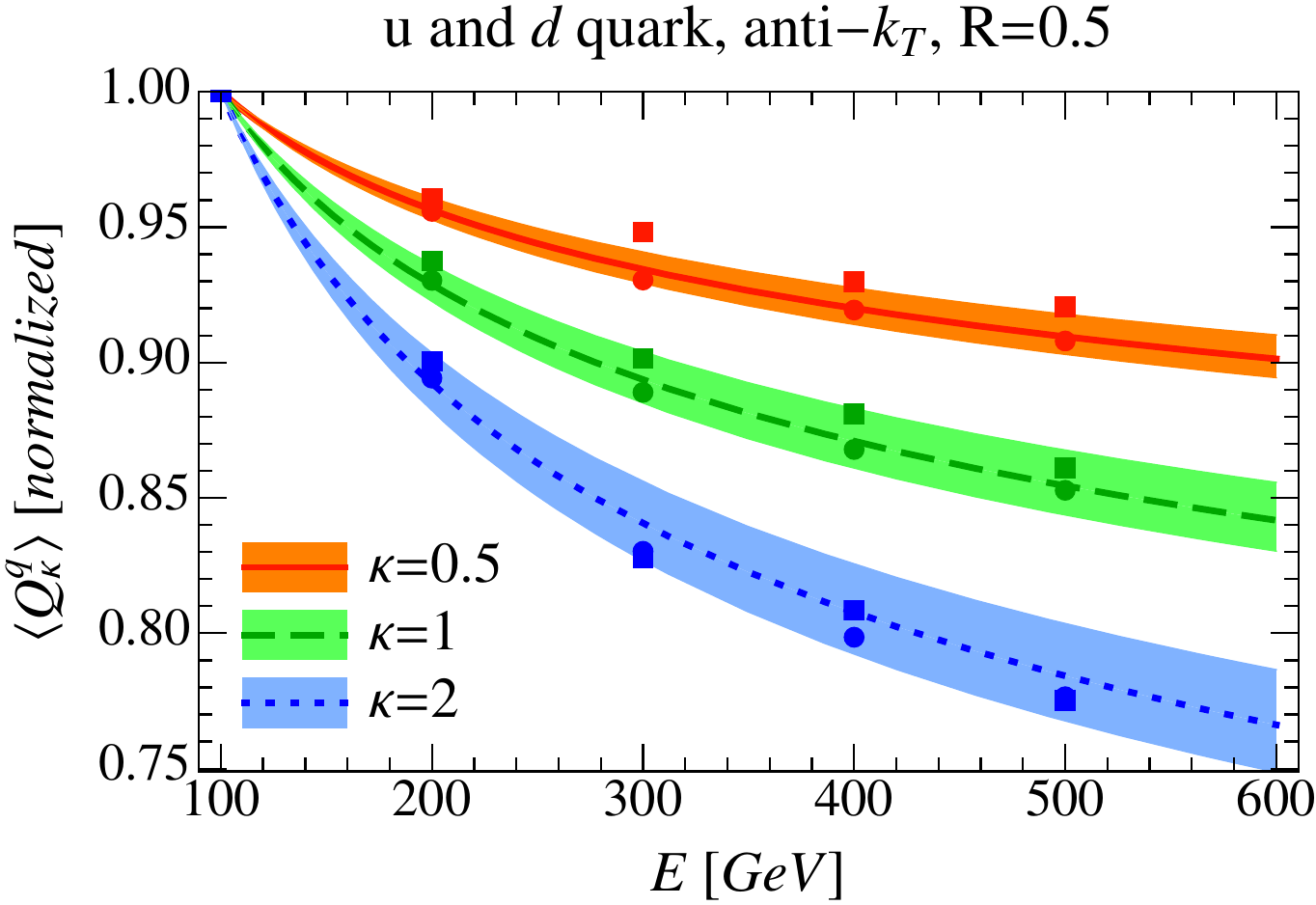} \quad
\includegraphics[width=0.46\textwidth]{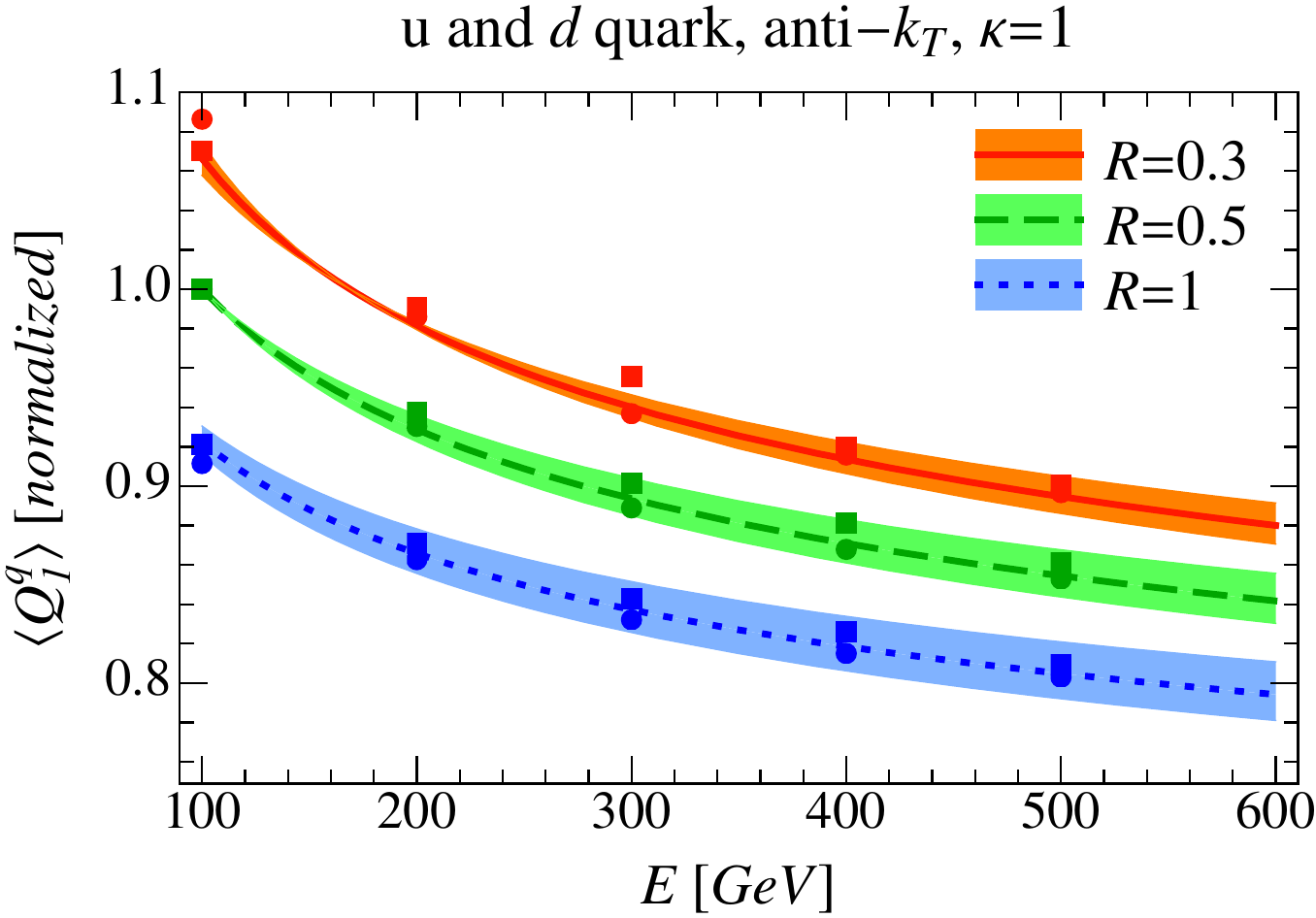} 
\caption{The average charge for an anti-$k_T$ quark jet is shown as function of the jet energy $E$ for various values of $\kappa$ and $R$. The \Pythia results for $d$ ($u$) quarks are shown as squares (circles). The plots are normalized to 1 at $E=100$ GeV and $R=0.5$, which removes the dependence on the nonperturbative input and thus the quark flavor.}
\label{fig:ave}
\end{figure*}

\begin{table*}
\begin{tabular}{|c|ccc|ccc|ccc|}
\hline
& \multicolumn{3}{c|}{$u$-quark} & \multicolumn{3}{c|}{$d$-quark} & \multicolumn{3}{c|}{$s$-quark} \\
$\kappa$ & \Pythia & DSS & AKK08 & \Pythia & DSS & AKK08 & \Pythia & DSS & AKK08 \\ \hline
0.5 & 0.271 & 0.237 & 0.221 & -0.162 & -0.184 &  -0.062 & -0.196 & -0.504 & -0.123 \\
1 & 0.144 & 0.122 & 0.134 & -0.078 & -0.088 & -0.046 & -0.108 & -0.214 & -0.064 \\
2 & 0.055 & 0.046 & 0.064 & -0.027 & -0.030 & -0.027 & -0.043 & -0.064 & -0.024 \\
\hline
\end{tabular}
\caption{Average charge of an $e^+e^-$ anti-$k_T$ jet with $E=100$ GeV and $R=0.5$. We did not include HKNS in this comparison due to its poor charge separation.}
\label{tab:qave}
\end{table*}
\begin{table*}
\begin{tabular}{|c|cccc|cccc|cccc|cccc|}
\hline
& \multicolumn{4}{c|}{$u$-quark} & \multicolumn{4}{c|}{$d$-quark} & \multicolumn{4}{c|}{$s$-quark} & \multicolumn{4}{c|}{gluon} \\
$\kappa$ & \Pythia & HKNS & DSS & AKK08 & \Pythia & HKNS & DSS & AKK08 & \Pythia & HKNS & DSS & AKK08 & \Pythia & HKNS & DSS & AKK08 \\ \hline
0.5 & 0.341 & 0.862 & 0.734 & - & 0.338 & 0.785 & 0.707 & - & 0.336 & 0.549 & 0.674 & - & 0.356 & 0.813 & 0.773 & - \\
1 & 0.242 & 0.383 & 0.333 & 0.373 & 0.236 & 0.339 & 0.313 & 0.340 & 0.237 & 0.225 & 0.314 & 0.298 & 0.200 & 0.314 & 0.300 & 0.381 \\
2 & 0.136 & 0.155 & 0.143 & 0.157 & 0.127 & 0.134 & 0.131 & 0.129 & 0.132 & 0.093 & 0.139 & 0.130 & 0.074 & 0.117 & 0.109 & 0.127 \\
\hline
\end{tabular}
\caption{Width of the jet charge distribution $\Ga_\kappa^i$ for an $e^+e^-$ anti-$k_T$ jet with $E=100$ GeV and $R=0.5$. The dihadron contribution is calculated assuming no correlations, using \eq{uncor}.}
\label{tab:qwidth}
\end{table*}

\section{Numerical Results}
\label{sec:results}

This section contains our numerical results for the average and width of the jet charge distribution. We first study the perturbative convergence of our results, followed by a detailed comparison with \Pythia. We conclude by discussing the optimal choice for $\kappa$.

\subsection{Perturbative Convergence}

We start by studying the perturbative convergence of our calculation. In \fig{aveconv} we show the results for the average charge of a $k_T$-like quark jet with $R=0.5$ and $\kappa=1$, as a function of the jet energy $E$. The curves are normalized to 1 at $E=100$ GeV, which removes the dependence on the nonperturbative parameter in \eq{Qave2}. At LO we do not include the NLO jet algorithm corrections, i.e.~we take $\widetilde \cJ_{ij} = 2(2\pi)^3 \de_{ij}$. As \fig{aveconv} shows, the NLO corrections reduce the average jet charge by a non-negligible amount.

The perturbative uncertainties are estimated by varying the renormalization scale $\mu$ up and down by a factor of 2. To keep the normalization point fixed, we simultaneously vary the scale in the normalization. We show uncertainty bands both with (darker) and without (lighter) this additional prescription in \fig{aveconv}. In all the following plots we will use this additional prescription, which keeps the normalization point fixed and leads to smaller uncertainties. However, since these uncertainty bands do not quite overlap, they may be a bit too optimistic. In addition, the prescription causes the NLO band to be only slightly narrower than the LO result. (Neither of these issues are present for the lighter uncertainty bands.)

In \fig{widconv} we study the convergence of $\langle (Q_\kappa^i)^2 \rangle$ for $i=q,g$, which enters in the width in \eq{Ga}. We can no longer completely remove the nonperturbative input by normalizing, because of the mixing between quarks and gluons. We therefore make an assumption for
\begin{align}
\rho = \frac{\langle (Q_\kappa^g)^2\rangle}{\langle (Q_\kappa^q)^2 \rangle} \text{ \ at \ } \mu_0=1 \text{ GeV}, 
\end{align}
which we for simplicity take equal for all five light quark flavors. 
The solid curves and uncertainty bands correspond to $\rho=1$ and the dotted curves in \fig{widconv} correspond to $\rho=2$. We find again that the convergence is reasonable. The mixing causes the width to reduce more slowly as function of $E$. (For quarks the effect of the mixing is stronger if $\rho$ is larger, whereas for gluons it is the opposite way around.)

\begin{figure}[b]
\centering
\includegraphics[width=0.48\textwidth]{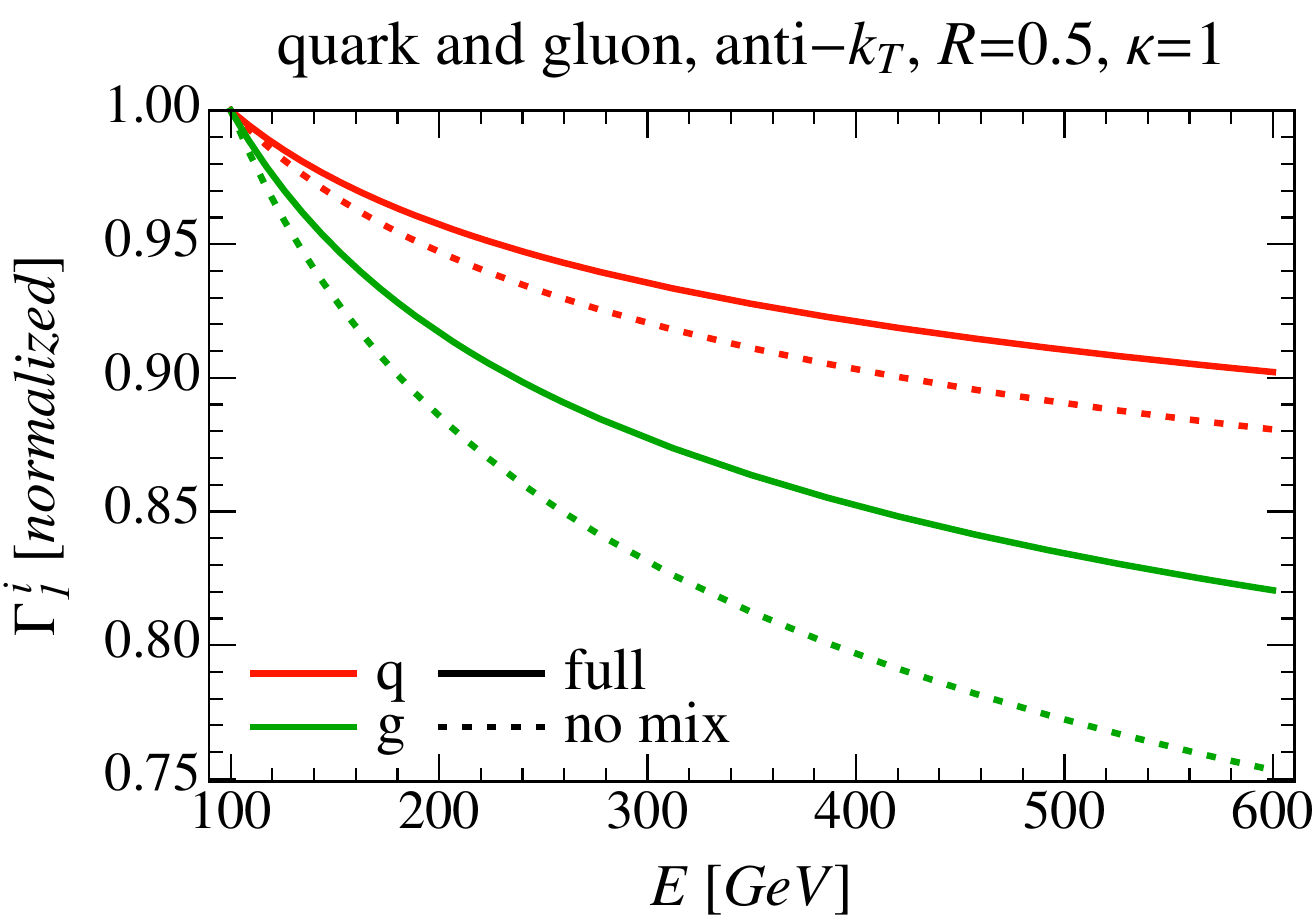}
\caption{The size of the quark/gluon mixing in the width of the jet charge distribution.}
\label{fig:mix}
\end{figure}

\begin{figure*}[t]
\centering
\includegraphics[width=0.48\textwidth]{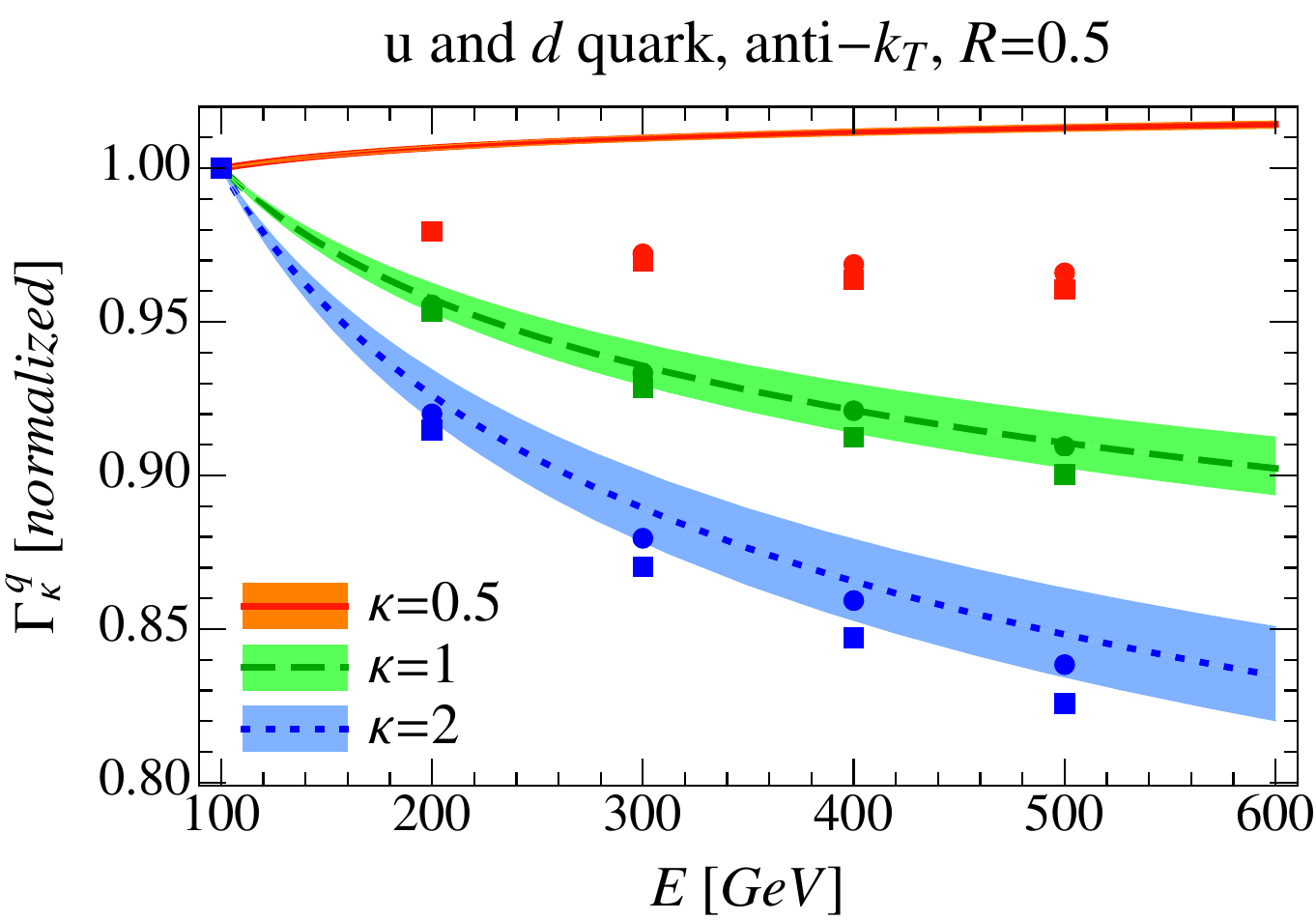} \quad
\includegraphics[width=0.48\textwidth]{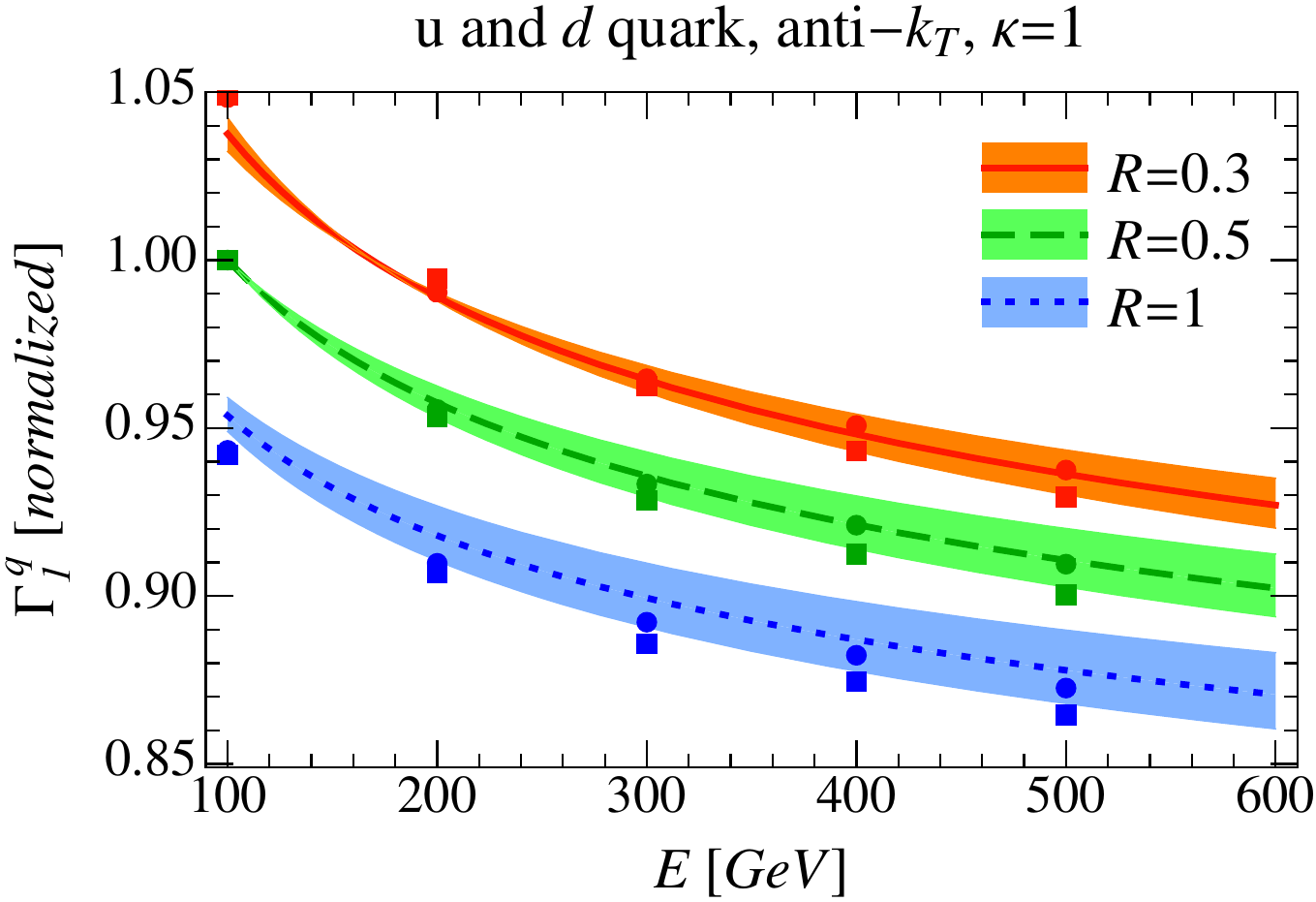} 
\caption{The width of the anti-$k_T$ quark jet charge distribution as a function of the jet energy $E$ for various values of $\kappa$ and $R$. The \Pythia results for $d$ ($u$) quarks are shown as squares (circles). The plots are normalized to 1 at $E=100$ GeV and $R=0.5$, since they are almost independent of the quark flavor.}
\label{fig:wid}
\end{figure*}

\begin{figure*}[t]
\centering
\includegraphics[width=0.48\textwidth]{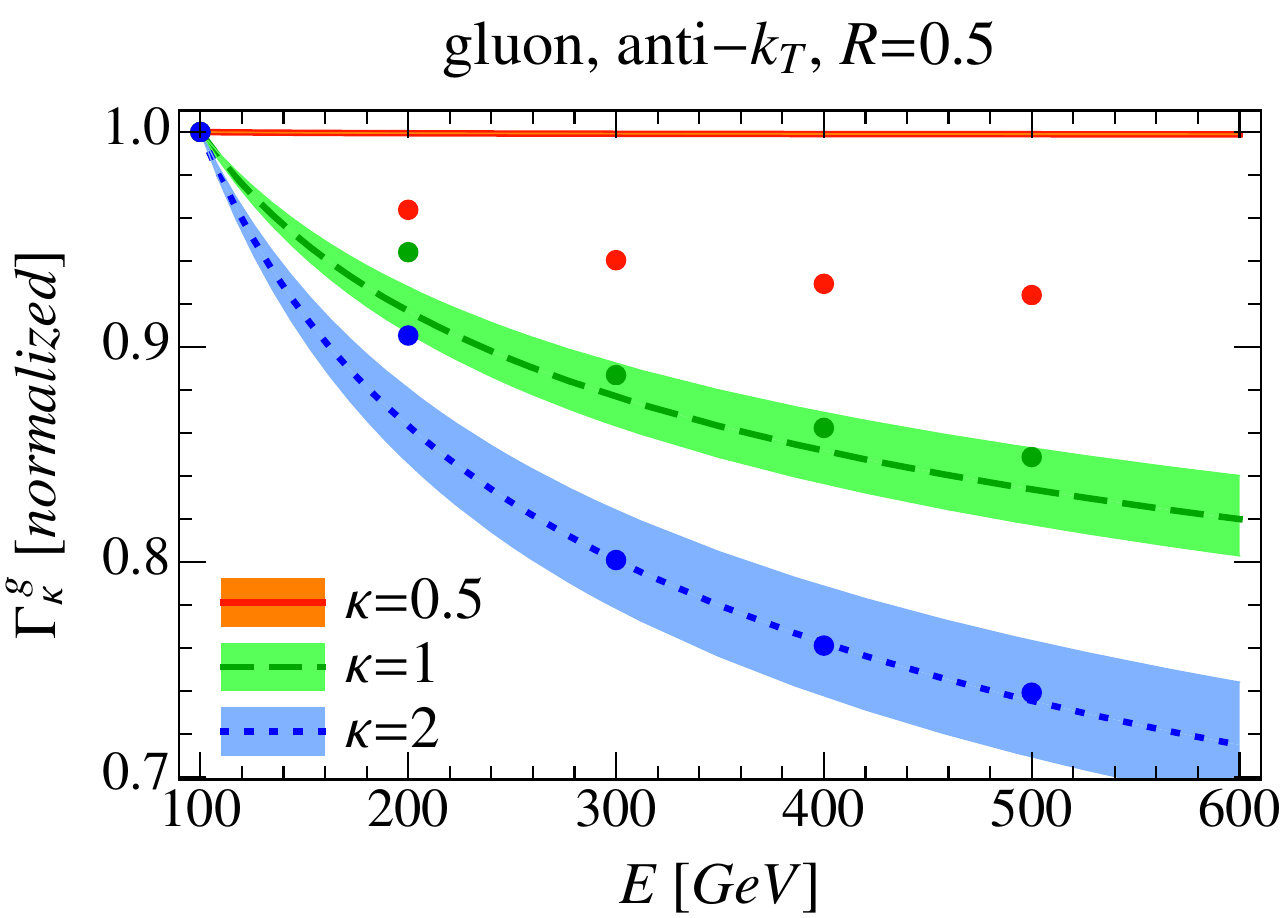} \quad
\includegraphics[width=0.48\textwidth]{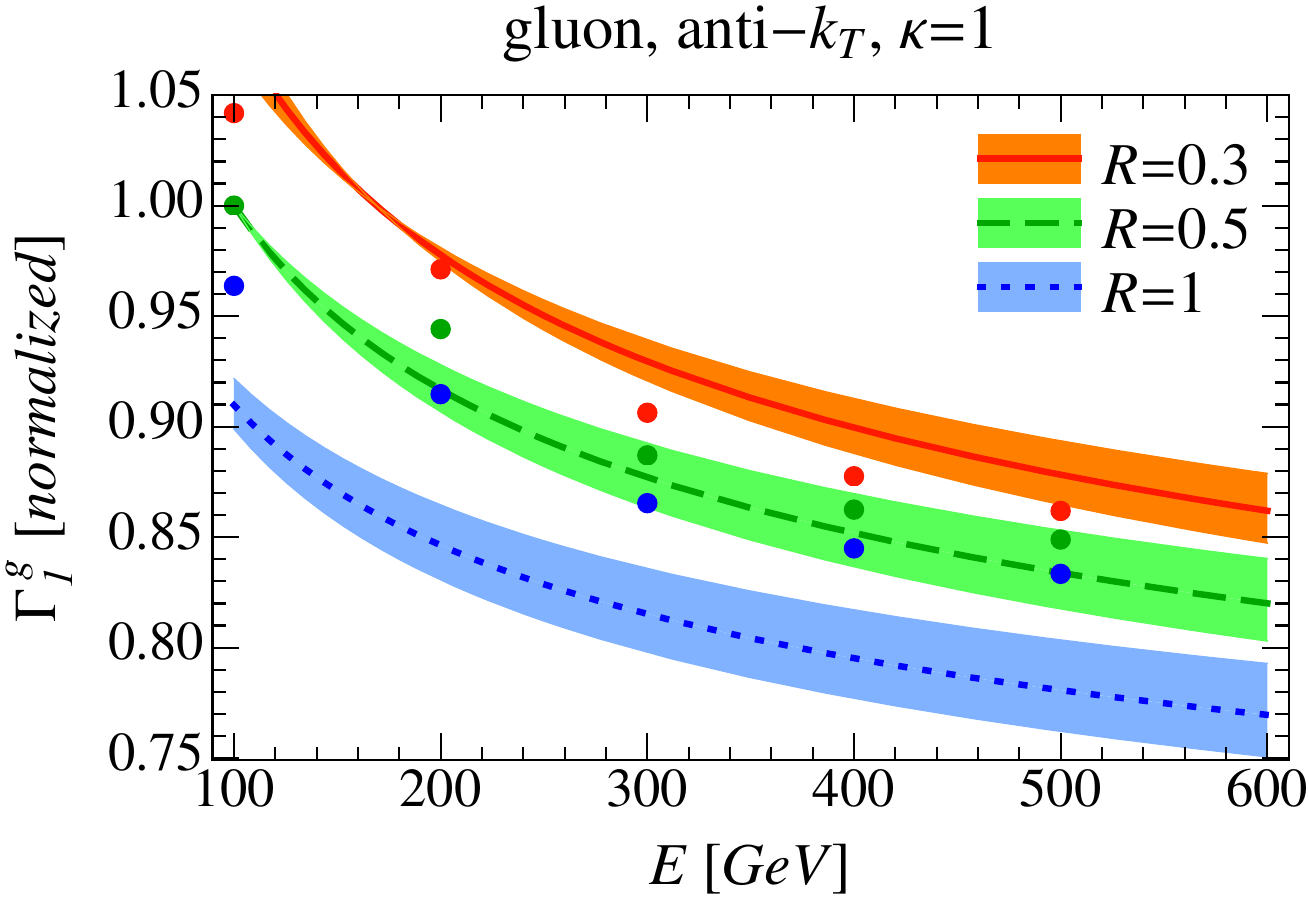} 
\caption{The width of the anti-$k_T$ gluon jet charge distribution as a function of the jet energy $E$ for various values of $\kappa$ and $R$. The \Pythia results are shown as circles, and the plots are normalized to 1 at $E=100$ GeV and $R=0.5$.}
\label{fig:widg}
\end{figure*}

\subsection{Comparison with \Pythia}

\subsubsection{Setup}

In this section we compare our calculation for the average and width of the jet charge distribution with \Pythia~\cite{Sjostrand:2006za,Sjostrand:2007gs}. \Pythia results for quark jets are obtained from the process $e^+e^- \to \ga/Z \to q\bar q$, where $\sqrt{s} = 2 E$ leads to jets of roughly the desired jet energy. The gluon jets are obtained from $pp \to gg$ by taking $\sqrt{s} = 2 E$ and requiring that the outgoing gluons have a minimum $p_T \sim E$.  We cluster the jets using the $e^+e^-$ version of the anti-$k_T$ algorithm in \FastJet~\cite{Cacciari:2011ma}, and only keep jets that have at least 95\% of the desired energy.

\subsubsection{Comparison using Fragmentation Functions}

First we compare the average jet charge, using the input from the HKNS~\cite{Hirai:2007cx}, DSS~\cite{deFlorian:2007aj,deFlorian:2007hc} and AKK08~\cite{Albino:2008fy} fragmentation function sets collected in \app{nonpert}, with the result obtained from \Pythia at $E=100$ GeV and $R=0.5$. As Table \ref{tab:qave} shows, \Pythia is in reasonably good agreement with the FF results, given the large uncertainties from the FFs, for which we take the spread between the FF sets as an estimate. The main reason for this spread is poor charge separation, as discussed in \app{nonpert}.

Since the knowledge of dihadron FFs is very limited, we need either input from data or further assumptions to make predictions for the width. For example, recently dihadron FFs have been calculated in the NJL model~\cite{Casey:2012hg}. Here we will assume that the two hadrons are uncorrelated. Since the evolution in \eq{DQQ_evo} and the matching in \eq{Qsig} generate correlations, we need to be more specific: we assume that the dihadron FFs are uncorrelated at the low scale $\mu_0=1$ GeV,
\begin{align}
  D_j^{h_1,h_2}(z_1,z_2,\mu_0) = D_j^{h_1}(z_1,\mu_0) D_j^{h_2}(z_2,\mu_0)
\,,\end{align}
which leads to 
\begin{align} \label{eq:uncor}
  \widetilde D_j^{QQ}(\kappa,\kappa,\mu_0) = [\widetilde D_j^Q(\kappa,\mu_0)]^2
\,.\end{align}
From the tables with nonperturbative parameters in \app{nonpert}, we then see that the dihadron contribution $\widetilde D_j^{QQ}$ is typically (much) smaller than the single hadron contribution $\widetilde D_j^{Q^2}$. In Table \ref{tab:qwidth} we show the width of the jet charge in \Pythia, as well as the width obtained using \eq{uncor}. For $\kappa=0.5$ and 1 the width obtained from \Pythia is consistently smaller than that obtained from FFs. Of course this could indicate significant dihadron correlation effects and not necessarily a problem with \Pythia. As the knowledge of (dihadron) FFs progresses, the above analysis could start putting constraints on parameters (or tunes) of \Pythia.

\subsubsection{Perturbative Comparison}

Next we compare the dependence on $E$ and $R$ in \Pythia with our calculation, for which we take \Pythia's values at $E=100$ GeV and $R=0.5$ as input. In the left panel of \fig{ave}, we compare the average jet charge as a function of the jet energy $E$ for $u$ and $d$ quark jets and $\kappa=0.5,1,2$. Since our calculation predicts the same shape for $u$ and $d$ quark jets, we normalize the plot to 1 at $E=100$ GeV. This also removes the dependence on the non-perturbative input describing the hadronization\footnote{We observe this in \Pythia as well: the average jet charge before and after hadronization is the same up to an overall factor.}. Our result agrees well with \Pythia. In the right panel of \fig{ave} we compare different values of $R$ for $\kappa=1$, finding again good agreement with \Pythia. 

We have investigated the size of the nonlinearities and mixing in the evolution and fixed-order corrections to the width of the jet charge in \eqs{Qsig}{DQQ_evo}. The effect of the nonlinearities is less than 1\% until you get down to energies of only a few GeV, so they are irrelevant. However, the mixing effect is quite significant. In \fig{mix} we show the (normalized) widths of the jet charge for quark and gluon jets, with and without the mixing contribution.

In \fig{wid}, we study the width of the quark jet charge distribution as a function of the jet energy. We again show normalized results, since these are almost the same for the different quark flavors. In the left panel we compare our calculation to \Pythia for several values of $\kappa$. For $\kappa=1,2$ our calculation agrees well with \Pythia, but for $\kappa=0.5$ there is a significant difference. 
The rise in our prediction for $\kappa=0.5$ is due to the large mixing contribution from gluons (which was not included in Ref.~\cite{Krohn:2012fg}). \Pythia's decrease for $\kappa=0.5$ involves a cancellation between the parton shower and the hadronization model, since before hadronization \Pythia also predicts a (relative) increase. At the same time the $\ord{\lambda^{2\kappa}}$ corrections due to soft radiation are the largest for small values of $\kappa$. (Since soft radiation does not affect the \emph{average} jet charge, this would not be in contradiction with the agreement seen in \fig{ave}.) At this point it is not clear if the discrepancy indicates a problem with \Pythia or our calculation.  In the right panel of \fig{wid} we perform the comparison for different values of $R$ with $\kappa=1$, which agrees well with the result obtained from \Pythia.

In \fig{widg} we study the width of the gluon jet charge. There is again good agreement for $\kappa=1$ and 2. For $\kappa=0.5$, we find that the gluon width barely changes due to the mixing with quarks, whereas \Pythia yields a distinctly decreasing width. As the right panel shows, the $R$ dependence in \Pythia exhibits the same general features as our result, but the effect is smaller. These points also suggest that the gluon jets in \Pythia do not depend on $E$ and $R$ solely through the combination $2E \tan(R/2)$ [as predicted by our calculation and observed for quark jets in \fig{var}].

Based on these comparisons it seems reasonable to use \Pythia for first jet charge studies. Precision studies will presumably lead to discrepancies with Pythia, though this can of course be alleviated by retuning. By contrast, our higher-order calculation can be systematically improved to match the experimental precision.

\subsection{Optimal Choice for $\kappa$}

\begin{table}[b]
 \begin{tabular}{|c|ccc|} 
 \hline
 $E$(GeV) & $u$ & $d$ & $s$ \\ \hline
 25 & 0.29 & 0.29 & 0.32 \\
 100 & 0.28 & 0.25 & 0.31 \\
 400 & 0.26 & 0.25 & 0.29 \\
 \hline
 \end{tabular}
 \caption{Theoretical optimal choice $\kappa_*$ obtained from \Pythia, for $e^+e^-$ anti-$k_T$ jets with $R=0.5$ and the indicated flavor and jet energy $E$.}
\label{tab:kstar}
\end{table}

\begin{figure}[t]
\centering
\includegraphics[width=0.48\textwidth]{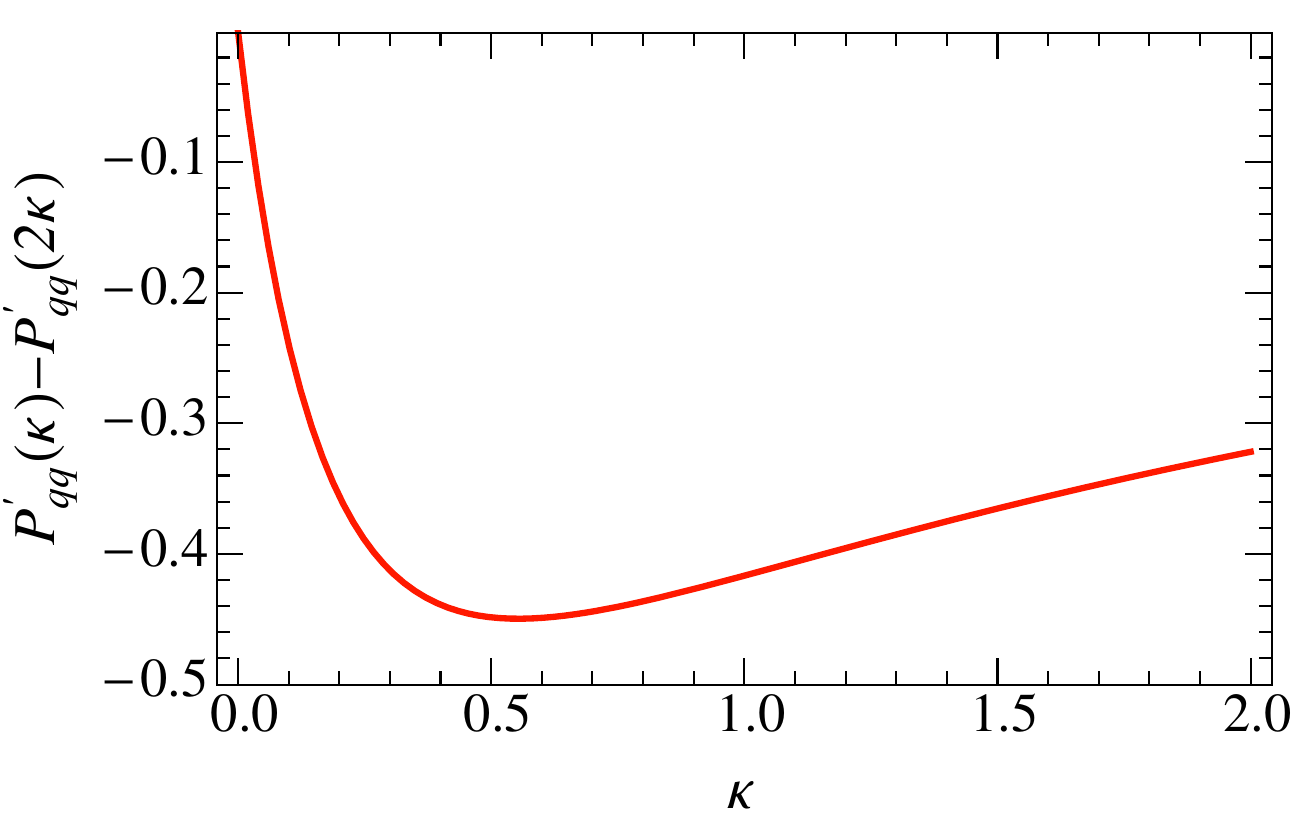}
\caption{Factor that controls the speed with which $\kappa_*$ reduces as $\mu \sim 2E \tan(R/2)$ increases. Below $\kappa \sim 0.55$ the reduction slows down.}
\label{fig:kstar}
\end{figure}

Theoretically, the optimal choice $\kappa_*$ is the value for $\kappa$ where the peak of the jet charge distribution is best separated from zero. More precisely, $\kappa_*$ is where
\begin{equation} 
  \eta(\kappa) = \frac{\langle Q_\kappa^q \rangle^2}{\langle (Q_\kappa^q)^2 \rangle}
\end{equation}
attains its maximum. Experimental considerations will of course affect this choice, since e.g.~smaller values of $\kappa$ increase the (unwanted) sensitivity to soft radiation from the initial state or other jets~\cite{Krohn:2012fg}.  Though the maximization of $\eta$ is a nonperturbative question, we can study how $\kappa_*$ depends on $\mu \sim 2E \tan(R/2)$. Neglecting NLO jet algorithm corrections and mixing with gluons,
\begin{equation}
   \frac{\df \eta}{\df \ln \mu} = \frac{\al_s(\mu)}{\pi} [2P_{qq}(\kappa) - P_{qq}(2\kappa)] \eta
\,.\end{equation}
This leads to
\begin{equation} \label{eq:kstar}
  \frac{\df\kappa_*}{\df \ln \mu} = - 2\,\frac{\al_s(\mu)}{\pi}  \frac{\eta(\kappa_*)}{\eta''(\kappa_*)} [P_{qq}'(\kappa_*) - P_{qq}'(2\kappa_*)]
\,,\end{equation}
so as we increase $E$ or $R$, $\kappa_*$ is reduced. The factor containing the splitting functions is plotted in \fig{kstar} and, interestingly\footnote{By contrast, $P_{gg}'(\kappa) - P_{gg}'(2\kappa) \sim -1/\kappa^2$, suggesting that any $z$-weighted property of gluon jets is quite diluted at high energies.}, has a minimum around $\kappa \sim 0.55$. Once $\kappa_*$ is below this value, its scale dependence will be reduced. In addition, $\al_s(\mu)$ reduces as $\mu$ increases. In table \ref{tab:kstar} we show some results for $\kappa_*$ obtained from \Pythia. Since the values of $\kappa_*$ are all well below 0.55, it is not surprising that there is little dependence on the jet energy. It should also be noted that $\eta$ is fairly flat in the vicinity of $\kappa_*$, so a somewhat different value of $\kappa_*$ may work almost as well.

\section{Conclusions}

We have presented in detail the calculation of the jet charge distribution. This takes a particularly simple form for the average and width of the jet charge distribution, which are the experimentally most relevant parameters. The nonperturbative coefficients that enter in the average and the width are related to moments of (dihadron) fragmentation functions. Since these currently still have large uncertainties the agreement with \Pythia is reasonably good. \Pythia tends to predict a width that is smaller than those obtained from fragmentation functions when dihadron correlations are neglected, but this could of course be due to such correlations. We also compared our perturbative calculation with the showering in \Pythia, which agree well for $\kappa \gtrsim 1$, suggesting that \Pythia suffices for initial studies of jet charge. The results in this paper can be systematically improved by including higher order corrections, power corrections or updated (dihadron) FFs. 

There are various choices that enter in the jet charge, such as the jet algorithm, $R$ and the weighting-power $\kappa$. Jets with a smaller $R$ retain a better jet charge signal, but also have an increased dependence on the jet algorithm. We studied the optimal choice $\kappa_*$ of $\kappa$ for quark jets and found that it reduces as the jet energy increases. However, this energy dependence slows down for $\kappa_*$ below 0.55, indicating that observables like jet charge can remain useful at high energies. 

We have also shown that our general calculation of jet charge can naturally be performed using a Monte Carlo-style approach. It is interesting that this is possible, given that standard Monte Carlo parton showers are limited to leading logarithmic order. We leave it to future work to investigate whether this can be extended to more general track-based jet observables that are also sensitive to soft radiation~\cite{Chang:2013rca}.

In Ref.~\cite{Krohn:2012fg} several potential applications of jet charge at the LHC were discussed. Here we confirmed through a detailed calculation that jet charge is theoretically under control. We therefore recommend a study of jet charge with LHC data as the natural next step.

\begin{acknowledgments}
We thank A.~Manohar, M.~Schwartz, G.~Stavenga and J.~Thaler for helpful discussions. We thank M.~Procura for feedback on this manuscript. This work was supported by DOE grant DE-FG02-90ER40546. 
\end{acknowledgments}

\appendix

\section{Perturbative Coefficients}
\label{app:pert}

\subsection{One-Loop Splitting Functions}

The one-loop splitting functions are \cite{Altarelli:1977zs}
\begin{align}
  P_{qq}(z) &= C_F \Big(\frac{1+z^2}{1-z}\Big)_+
  = C_F \Big[ \frac{1+z^2}{(1-z)}_+ \!\!+ \frac{3}{2}\, \de(1-z) \Big]
  \,, \nn \\
  P_{gq}(z) &= C_F\, \frac{1+(1-z)^2}{z}
  \,, \nn \\
  P_{gg}(z)
  &= 2 C_A \Bigl[\frac{z}{(1\!-\!z)}_+ \!\!+ \frac{1\!-\!z}{z} +  z(1\!-\!z)\Bigr]
   + \frac{1}{2} \beta_0\, \delta(1\!-\!z)
  \,,\nn\\
  P_{qg}(z) &= T_F [z^2+(1-z)^2]
  \,,
\end{align}
where $\beta_0 = (11 C_A - 4 n_f T_F)/3$, is the lowest order coefficient of the QCD $\beta$-function. In moment space~\cite{Gross:1974cs}
\begin{align} 
  \widetilde P_{qq}(\kappa) &= C_F \Big[-2 H(\kappa+2)+ \frac{1}{\kappa+1} + \frac{1}{\kappa+2}+ \frac{3}{2}\Big]
  \,, \nn \\
  \widetilde P_{gq}(\kappa) &= C_F\, \frac{\kappa^2+3\kappa+4}{\kappa^3+3\kappa^2+2\kappa}
  \,, \nn \\
  \widetilde P_{gg}(\kappa)
  &= C_A \Big[-2H(\kappa+3) + \frac{4\kappa^2+6\kappa+4}{\kappa^3+3\kappa^2+2\kappa}\Big]  + \frac{1}{2} \beta_0
  \,,\nn\\
  \widetilde P_{qg}(\kappa) &= T_F\, \frac{\kappa^2+3\kappa+4}{\kappa^3+6\kappa^2+11\kappa+6}
  \,,\nn\\
  \widehat P_{qg}(\kappa) &= T_F\, \frac{2\Ga(\kappa+1) \Ga(\kappa+3)}{\Ga(2\kappa+4)}
  \,,
\end{align}
where $H$ is the harmonic number function. The nonlinear contribution $\widehat P_{qg}$ decreases exponentially for large $\kappa$,
\begin{align}
 \widehat P_{qg}(\kappa) & =
 T_F\, \frac{\sqrt{\pi} e^{-(2 \ln 2) \kappa}}{4\sqrt{\kappa}} \Big[1 + \ORd{\frac{1}{\kappa}}\Big]
\,.\end{align}

\subsection{NLO corrections for $k_T$-type jets}
\label{app:kt}

\begin{table}[b]
 \begin{tabular}{|c|ccccl|} 
 \hline
 $\kappa$ & $\widetilde \cJ_{qq}$ & $\widetilde \cJ_{qg}$ & $\widetilde \cJ_{gg}$ & $\widetilde \cJ_{gq}$ & $\widehat \cJ_{gq}$ \\ \hline
0.5 & 1.07 & -8.49 & -6.83 & -0.81 & -0.39 \\
1 & 2.10 & -2.56 & 0.02 & -0.64 & -0.14 \\
2 & 3.65 & -1.01 & 3.01 & -0.50 & -0.023 \\
3 & 4.85 & -0.67 & 4.51 & -0.43 & -0.0045 \\
4 & 5.86 & -0.53 & 5.64 & -0.38 & -0.00094 \\
 \hline
 \end{tabular}
 \caption{Numerical results for the moments of the one-loop $\cJ_{ij}^\fin$ (does not include the overall $\al_s/\pi$ and color factor).}
 \label{tab:J_k}
\end{table}

\begin{table*}[t]
\begin{tabular}{|c|ccc|ccc|ccc|}
\hline
 & \multicolumn{3}{c|}{HKNS NLO} & \multicolumn{3}{c|}{DSS NLO} & \multicolumn{3}{c|}{AKK08} \\
$\kappa$ & $\widetilde D_u^Q$ & $\widetilde D_d^Q$ & $\widetilde D_s^Q$ & $\widetilde D_u^Q$ & $\widetilde D_d^Q$ & $\widetilde D_s^Q$ & $\widetilde D_u^Q$ & $\widetilde D_d^Q$ & $\widetilde D_s^Q$ \\ \hline
0.5 & 1.207 & -0.807 & -0.073 & 0.302 & -0.235 & -0.642 & 0.279 & -0.079 & -0.156 \\
1 & 0.420 & -0.279 & -0.062 & 0.184 & -0.132 & -0.323 & 0.199 & -0.068 & -0.095 \\
2 & 0.135 & -0.089 & -0.039 & 0.087 & -0.057 & -0.121 & 0.120 & -0.051 & -0.045 \\
\hline
\end{tabular}
\caption{Nonperturbative parameters $\widetilde D_q^Q(\kappa,\mu=1\, \text{GeV})$ for the average charge of a quark jet.}
\label{tab:dq}
\end{table*}

\begin{table*}[t]
\begin{tabular}{|c|cccc|cccc|cccc|}
\hline
 & \multicolumn{4}{c|}{HKNS NLO} & \multicolumn{4}{c|}{DSS NLO} & \multicolumn{4}{c|}{AKK08} \\
$\kappa$ & $\widetilde D_u^{Q^2}$ & $\widetilde D_d^{Q^2}$ & $\widetilde D_s^{Q^2}$ & $\widetilde D_g^{Q^2}$ & $\widetilde D_u^{Q^2}$ & $\widetilde D_d^{Q^2}$ & $\widetilde D_s^{Q^2}$ & $\widetilde D_g^{Q^2}$ & $\widetilde D_u^{Q^2}$ & $\widetilde D_d^{Q^2}$ & $\widetilde D_s^{Q^2}$ & $\widetilde D_g^{Q^2}$ \\ \hline
1 & 0.676 & 0.540 & 0.113 & 0.680 & 0.498 & 0.442 & 0.347 & 0.620 & - & - & - & - \\
2 & 0.206 & 0.161 & 0.051 & 0.291 & 0.165 & 0.143 & 0.130 & 0.260 & 0.188 & 0.150 & 0.099 & 0.447 \\
4 & 0.050 & 0.038 & 0.016 & 0.108 & 0.045 & 0.038 & 0.040 & 0.090 & 0.050 & 0.034 & 0.035 & 0.120 \\
\hline
\end{tabular}
\qquad
\begin{tabular}{|c|cc|cc|cc|}
\hline
 & \multicolumn{2}{c|}{HKNS NLO} & \multicolumn{2}{c|}{DSS NLO} & \multicolumn{2}{c|}{AKK08} \\
$\kappa$ & $\widetilde D_c^{Q^2}$ & $\widetilde D_b^{Q^2}$ & $\widetilde D_c^{Q^2}$ & $\widetilde D_b^{Q^2}$ & $\widetilde D_c^{Q^2}$ & $\widetilde D_b^{Q^2}$ \\ \hline
1 & 0.510 & 0.607 & 0.618 & 0.639 & - & 0.459\\
2 & 0.108 & 0.071 & 0.145 & 0.081 & 0.086 & 0.060\\
4 & 0.013 & 0.004 & 0.022 & 0.008 & 0.011 & 0.004 \\
\hline
\end{tabular}
\caption{Nonperturbative parameters $\widetilde D_i^{Q^2}(\kappa,\mu)$ that contribute to the width of jet charge distributions. For $i=u,d,s,g$ the scale $\mu=1$ GeV and for $i=c,b$ it is $\mu = m_{c,b}$. For AKK08 most of the $\kappa=1$ moments are divergent, as denoted by ``-".}
\label{tab:dq2}
\end{table*}

The NLO matching coefficients $\cJ_{ij}$ for the $e^+e^-$ version of $k_T$-like jet algorithms are given below, and we will also discuss its straightforward extension to $pp$ collisions. At one-loop, where you have at most two partons, there is no distinction between the various $e^+e^-$ $k_T$-like jet algorithms. The jet restriction is simply $\theta \leq R$, where $\theta$ is the angle between the two partons. This translates into
 \begin{equation}
   s \leq 4z(1-z)E^2 \tan^2 (R/2)
\,,\end{equation}
where $s$ is the invariant mass of the jet. For $k_T$-like algorithms for $pp$ collisions the corresponding jet restriction is $\sqrt{(\Delta \eta)^2 + (\Delta \phi)^2} \leq R$, where $\Delta \eta$ and $\Delta \phi$ are the difference in (pseudo)rapidity and azimuthal angle between the two partons. This is a more complicated restriction, as it is not rotationally symmetric around the jet axis. However, for narrow jets ($R \ll 1$) it simplifies to~\cite{Mukherjee:2012uz} 
 \begin{equation}
   s \leq 4z(1-z)p_T^2 \tan^2 (R/2)
\,.\end{equation}
Our $e^+e^-$ results can thus directly be extended to $pp$ collisions by simply replacing the jet energy $E$ by the transverse momentum $p_T$ of the jet.

Using the bare results in Ref.~\cite{Jain:2011xz}, we readily obtain, in the $\overline{\text{MS}}$ scheme,
\begin{align} \label{eq:Jij}
   \frac{\cJ_{qq}(E,R,z,\mu)}{2(2\pi)^3} &= \de(1-z) + \frac{\al_s(\mu)}{\pi} \bigg\{C_F L^2\, \de(1-z)  
   \nn \\ & \quad  
   \!+\! \Big[P_{qq}(z) \!-\! \frac{3}{2} C_F\, \de(1\!-\!z)\Big] L \!+\! C_F \cJ_{qq}^\fin(z) \bigg\}
   \,,\nn \\
   \frac{\cJ_{qg}(E,R,z,\mu)}{2(2\pi)^3} &= \frac{\al_s(\mu)}{\pi} \big[P_{gq}(z) L + C_F \cJ_{qg}^\fin(z) \big]
   \,,\nn \\
   \frac{\cJ_{gg}(E,R,z,\mu)}{2(2\pi)^3} &= \de(1-z) + \frac{\al_s(\mu)}{\pi} \bigg\{C_A L^2\, \de(1-z) 
   \nn \\ & \quad
   \!+\! \Big[P_{gg}(z) \!-\! \frac{1}{2}\bt_0\, \de(1\!-\!z) \Big] L \!+\! C_A \cJ_{gg}^\fin(z) \bigg\}
   \,,\nn \\
   \frac{\cJ_{gq}(E,R,z,\mu)}{2(2\pi)^3} &= \frac{\al_s(\mu)}{\pi} \big[P_{qg}(z) L +  T_F \cJ_{gq}^\fin(z) \big]
\,.\end{align}
Here $L = \ln [2 E\tan (R/2)/\mu]$ and
\begin{align} \label{eq:Jijhat}
  \cJ_{qq}^\fin(z) &= 
  2 z \Big(\frac{\ln(1-z)}{1-z}\Big)_{\!+}  +(1-z) \ln (1-z)
  \nn \\ & \quad
  + \frac{1+z^2}{1-z} \ln z +  \frac{1}{2} (1- z)  - \frac{\pi^2}{24}\, \de(1-z)
  \,, \nn \\
  \cJ_{qg}^\fin(z) &= \frac{1+(1-z)^2}{z} \ln [z(1-z)] +  \frac{z}{2} 
  \,, \nn \\
  \cJ_{gg}^\fin(z) &= 2 z \Big(\frac{\ln(1-z)}{1-z}\Big)_{\!+} \!\!+ \frac{2(1+z^2)(1-z)}{z} \ln(1-z) \nn \\ & \quad
    +\frac{2(1-z+z^2)^2}{z(1-z)} \ln z - \frac{\pi^2}{24}\, \de(1-z)
  \,, \nn \\  
  \cJ_{gq}^\fin(z) &=  [z^2+(1-z)^2] \ln[z(1-z)]+ z(1-z)
\,.\end{align}
Anti-quarks have the same coefficients as quarks, and $\cJ_{q\bar q}$ and $\cJ_{qq'}$ only start at two-loop order. We have checked these results using the sum rules in Ref.~\cite{Procura:2011aq}. Since we evaluate the matching coefficients at $\mu \sim 2E \tan(R/2)$, where $L \sim 0$, the one-loop contribution essentially comes from $\cJ_{ij}^\fin$. We give numerical values for its moments in Table \ref{tab:J_k}. Note that the nonlinear contribution $\widehat \cJ_{gq}$ becomes negligibly small for large $\kappa$.

The jet functions for $k_T$-like jets are~\cite{Ellis:2010rwa}
\begin{align} \label{eq:J}
  J_q(E,R,\mu) &= 1+\frac{\al_s(\mu) C_F}{\pi} \Big(L^2 - \frac{3}{2} L + \frac{13}{4} - \frac{3\pi^2}{8} \Big)
 \,, \nn \\
  J_g(E,R,\mu) &= 1+\frac{\al_s(\mu)}{\pi} \Big[C_A L^2 - \frac{\beta_0}{2} L +
  \nn \\
  & \quad C_A\Big(\frac{5}{24} - \frac{3\pi^2}{8}\Big)+ \frac{23}{24} \beta_0 \Big]
\,.\end{align}

\section{Nonperturbative Coefficients from Fragmentation Functions}
\label{app:nonpert}

In Table \ref{tab:dq} we show the nonperturbative parameters $\widetilde D_q^Q(\kappa,\mu=1\, \text{GeV})$ for the average charge of a quark jet, using the HKNS~\cite{Hirai:2007cx}, DSS~\cite{deFlorian:2007aj,deFlorian:2007hc} and AKK08~\cite{Albino:2008fy} fragmentation function sets at NLO. The large differences between the various FF sets are mainly due to poor charge separation. This is because we need the charge-separated combination $D_q^h - D_q^{\bar h} = D_q^h - D_{\bar q}^h$, whereas a lot of the data is $e^+e^-\to hX$ which only gives access to $D_q^h + D_{\bar q}^h$. In particular, HKNS only uses $e^+e^-$ data in their analysis, so their quark/anti-quark separation relies crucially on assumptions. The large difference for the $s$-quark in DSS compared to the other FF sets is due to semi-inclusive DIS data that only they include~\cite{deFlorian:2007aj}.

The results for $\widetilde D_i^{Q^2}$, which is the contribution from single hadron FFs to the width of the jet charge distribution, is shown in Table \ref{tab:dq2}. Several of the $\kappa=1$ moments for AKK08 are divergent (at $z=0$) and denoted by a ``-". Since $\widetilde D_q^{Q^2}$ essentially depends on the combination $D_q^h + D_{\bar q}^h$, the agreement between the different sets is much better. The gluon FFs are not as well known as the quark FFs, as is clear from the differences between the FF sets. We have also included the heavy quark flavors because they contribute through RG mixing.

\bibliography{jet_charge}

\end{document}